\documentclass[12pt,draftclsnofoot,onecolumn,journal,comsoc]{IEEEtran}

\ifCLASSOPTIONcompsoc
  \usepackage[nocompress]{cite}
\else
  \usepackage{cite}
\fi
\usepackage{amsthm}
\usepackage{amssymb}
\usepackage{amsmath}
\usepackage[T1]{fontenc}
\usepackage[utf8]{inputenc}
\usepackage[english]{babel}
\usepackage{graphicx}
\usepackage[short]{optidef} 

\ifCLASSOPTIONcompsoc
  \usepackage[caption=false,font=footnotesize,labelfont=sf,textfont=sf]{subfig}
\else
  \usepackage[caption=false,font=footnotesize]{subfig}
\fi
\usepackage{tikz}
\usetikzlibrary{datavisualization.formats.functions, backgrounds}
\usepackage{pgfplots}
\pgfplotsset{compat=1.14} 

\newtheorem{definition}{Definition}
\newtheorem{lemma}{Lemma}
\newtheorem{theorem}{Theorem}
\newtheorem{corollary}{Corollary}

\newtheorem{conjecture}{Conjecture}

\theoremstyle{remark}

\newcommand{\R}{\mathbb{R}}

\newcommand{\zero}{\mathbf{0}}

\newcommand{\prob}{\mathcal{P}}

\DeclareMathOperator*{\var}{Var}

\DeclareMathOperator*{\out}{out}
\DeclareMathOperator*{\off}{off}
\DeclareMathOperator*{\on}{on}
\DeclareMathOperator*{\singb}{SB}
\DeclareMathOperator*{\unc}{unc}
\DeclareMathOperator*{\cod}{cod}

\begin{document}
%
\title{Resilient Design of 5G Mobile-Edge Computing Over Intermittent mmWave Links}
%
%
%
%

\author{Nicola~di~Pietro,~
        Mattia~Merluzzi,~\IEEEmembership{Student Member,~IEEE,}
        Emilio~Calvanese~Strinati,~\IEEEmembership{Member,~IEEE,}
        and~Sergio~Barbarossa,~\IEEEmembership{Fellow,~IEEE}
\IEEEcompsocitemizethanks{\IEEEcompsocthanksitem N.~di Pietro and E.~Calvanese Strinati are with CEA-Leti, 17 rue des Martyrs, 38000 Grenoble, France.
\protect\\
E-mail: nicola.dipietro@cea.fr, emilio.calvanese-strinati@cea.fr
\IEEEcompsocthanksitem M.~Merluzzi and S.~Barbarossa are with the Department of Information Engineering, Electronics, and Telecommunications of Sapienza University, via Eudossiana 18, 00184 Roma, Italy.
\protect\\
E-mail: mattia.merluzzi@uniroma1.it, sergio.barbarossa@uniroma1.it
}
}

%
%


\markboth{}%
{}

%



\IEEEtitleabstractindextext{%
\begin{abstract}
Two enablers of the 5th Generation (5G) of mobile communication systems are the high data rates achievable with millimeter-wave radio signals and the cloudification of the network's mobile edge, made possible also by Multi-access Edge Computing (MEC). In 5G networks, user devices may exploit the high capacity of their mobile connection and the computing capabilities of the edge cloud to \textit{offload} computational tasks to MEC servers, which run applications on devices’ behalf. This paper investigates new methods to perform power- and latency-constrained offloading. First, aiming to minimize user devices' transmit power, the opportunity to exploit concurrent communication links between the device and the edge cloud is studied. The optimal number of channels for simultaneous transmission is characterized in a deterministic and a probabilistic scenario. Subsequently, blocking events that obstruct millimeter-wave channels making them “intermittent” are considered. Resource overprovisioning and error-correcting codes against asymmetric block erasures are proposed to jointly contrast blocking and exploit multi-link communications' diversity. The asymmetric block-erasure channel is characterized by a study of its outage probability. The analysis is performed in a framework that yields closed-form expressions. These, together with corroborating numerical results, are intended to provide reference points and bounds to optimal performance in practical applications.
\end{abstract}

}

\maketitle

\IEEEdisplaynontitleabstractindextext

%
\IEEEpeerreviewmaketitle

\section{Introduction}
\label{sec:introduction}

%
%
%
%
\IEEEPARstart{M}{obile}
data traffic is facing an impressive growth and it is foreseen to reach about 100 Exabytes per month in 2023 \cite{Ericsson18}. The conception of the fifth generation of mobile systems (5G) had its first achievement with the completion of 3GPP's Release 15~\cite{3GPP} and it is now at its second phase. The first goal of 5G is a 1000-fold enhancement of the system area capacity, achievable by the exploitation of massive MIMO, a dense deployment of small cell base stations, and larger bandwidths \cite{Andrews2014}. These three solutions are possible also thanks to the introduction of millimeter wave (mmWave) communications for radio access in the mobile environment \cite{Rap13,Sak17}. However, 5G  networks are foreseen to go beyond the enhancement of the physical layer, aiming to enable several new services for different sectors (\textit{verticals}), such as Internet of Things (IoT), industry 4.0, autonomous vehicles, remote surgery, etc. All of these services have such different requirements in terms of latency, reliability, etc., that a flexible design of the network is needed to fulfill each of them. This is possible thanks to network slicing and network function virtualization \cite{Cha17}. Due to the stringent latency requirements of new applications, such as augmented or virtual reality, there is the need for deploying computation and storage resources close to the end users, in order to reduce the time to reach the cloud and, at the same time, alleviate the load on backhaul networks. A key role in this framework will be played by Multi-access Edge Computing (MEC) \cite{ETSI-MEC}, a technology standardized by ETSI whose aim is to bring computation capabilities to the edge of the network, either in the radio access network or at an aggregation point. The merge of MEC and mmWave communications is the idea behind the Euro-Japanese project 5G-MiEdge to enable the 5G ecosystem~\cite{Miedge}. It is believed that these two technologies can compensate each other's drawbacks and benefit from each other's potentials to provide the services promised by the 5G vision. Indeed, mmWave can enable a fast access to MEC resources to provide low-latency services, whereas the computation resources of MEC can be used to orchestrate such a complex radio access network in terms of interference management, beamforming optimization, etc.

In this paper, we focus on \emph{computation offloading} \cite{Mach17}, an application enabled by MEC by which the execution of computationally heavy applications can be transferred from a user device to a server, which in case of MEC is called Mobile Edge Host (MEH). Offloading applications is convenient for different reasons, e.g.~reducing the energy consumption at the user's side and enabling resource-poor devices to run sophisticated applications. Computation offloading is composed by: a first communication phase, during which the bits necessary to execute the application are transmitted to a MEC Access Point (AP) and then to the MEH; a computation phase, during which the bits are processed by the MEH; and a second communication phase, in which the result is transmitted back to the end user. Some applications require strict latency constraints, so that mmWave coupled with MEC are needed to enable their offloading. Several works~\cite{Bar14,Sar18,Lin16} investigate the problem of resource allocation for computation offloading, showing the convenience of jointly optimizing radio and computation resources. For a comprehensive survey on MEC and computation offloading, the reader may see~\cite{Mao_chang17}. 

The major drawback of mmWave communications is their vulnerability to blocking events due to obstacles or beam collisions \cite{Abouelseoud2013,Singh2011}. When a blocking event occurs, the attenuation is so high that the communication is interrupted. The passage of obstacles between the transmitter and the receiver causes a certain ``intermittency'' of the channel that may lead to losses of information. Then, the latter has to be either retransmitted or recovered via suitably designed error-correcting coding schemes. During the communication phases of the computation offloading procedure, there may be no time to retransmit the information bits, since blocking events can last much longer than the maximum allowed offloading latency; moreover, the retransmission can lead to a high additional power consumption. Many offloading problems are formulated as the minimization of the power consumption at the mobile side and blocking events can be detrimental in this direction. Then, different countermeasures can be taken to deal with blocking events \cite{Barbarossa2017,BarbarossaCeci2017,Oguma2015}. In~\cite{BarbarossaCeci2017} and \cite{Barbarossa2017}, we investigated solutions based on multi-link communications and overprovisioning of radio and computation resources, taking into account an a priori knowledge (estimation) of the blocking probabilities. In these works, we formulated the problem as the minimization of the power consumption to guarantee an average bit rate above a certain threshold. In \cite{Oguma2015}, the authors perform a proactive AP selection based on prediction of human blocking events. In \cite{Gio16}, uplink channel measurements are used for the selection of the best AP, and to select a new AP in case a blocking event occurs. In \cite{Ohmann}, the authors investigate the problem of achieving high availability in wireless networks exploiting an optimal number of Rayleigh fading links. Some potential architectural options for multi-connectivity are described in \cite{Rav16}. Multi-connectivity can refer, in general, to the access to different Radio Access Technologies (RAT), such as Long Term Evolution (LTE) and 5G, or to the access to multiple interfaces of the same RAT. One possible architecture is the common Medium Access Control (MAC) solution, in which the multi-connectivity legs share the Packet Data Convergence Protocol (PDCP), Radio Link Control (RLC), and MAC layers and the physical layer is separated in different Remote Radio Heads (RRHs). In this paper, we exploit a similar architecture, in which multiple APs are employed in the uplink direction to simultaneously counteract blocking events and to reduce the transmitter's power consumption. 

\subsection{Our Contributions}
In the context of computation offloading to edge clouds, this paper proposes new solutions for reducing the uplink transmit power for end users under delay constraints and simultaneously contrasting the blocking events typical of mmWave communications. Our analyses 
allows to derive mathematically clean results and closed-form expressions. Globally, the goal of this paper is to prove the validity of the proposed schemes in a theoretical framework that will serve as a reference point and as a bound to optimal performance in practical applications.

After formally recalling  the problem of computation offloading and fixing some notation in Section~\ref{sec:offloading}, in Section~\ref{sec:multi_link} we treat the problem of transmit power minimization via simultaneous multi-link offloading. More precisely, we consider the possibility that a User Equipment (UE) offloads an application by splitting the total information into different blocks to be concurrently transmitted to different APs of the edge cloud. In a scenario without blocking and with only line-of-sight communication paths, we suppose that the modulation and coding schemes over each link are chosen to achieve the maximum transmission rate and that the mmWave beams employed for communications are narrow enough to make negligible the risk of beam collisions. Under these hypotheses, we fully characterize in Theorem~\ref{thm:N_link} and Corollary~\ref{cor:N_star} the multi-link strategy that achieves the same spectral efficiency of the single-link case while minimizing the power consumption. This approach is conceptually motivated by the idea that it may be pointless to deliver a service in the absolutely quickest possible manner; instead, it is preferable to fix a given satisfactory service latency constraint and focus on spending the least energy possible to meet it. Furthermore, in Section~\ref{sec:opt_link_distribution}, we analyze the probability distribution of the optimal number of links (in the sense defined in Section~\ref{sec:multi_link}), when the AP deployment follows a homogeneous Poisson point process. Interestingly, we manage to express a closed formula for this distribution. We conclude the section by describing how this expression can be exploited to fix an AP deployment density that guarantees power minimization on the end users' side with high probability. 

In the second part of the paper, we add to our analysis the possibility that mmWave communication links undergo blocking events. In Section~\ref{sec:multilink_with_blocking}, we recall the ``overprovisioning'' method, initially introduced in~\cite{BarbarossaCeci2017} and~\cite{Barbarossa2017}, and show how it can be successfully combined with multi-link offloading. This solution is useful to contrast short-term blocking events that occur after the beginning of the offloading procedure and whose duration is much shorter than the offloading time. This kind of blocking can be seen as momentary and brief channel availability interruptions with respect to the duration of the application. They can be fought by ``catching up'' with the offloading procedure as soon as they are over or, analogously, by performing the whole procedure at a higher average information transmission rate to compensate the time loss that they cause. In Section~\ref{sec:coding}, instead, we face the problem of long-term blocking events that start after the beginning of the offloading procedure and last as much as or more than the maximum tolerable latency. In this case, analogous solutions to the overprovisioning of Section~\ref{sec:multilink_with_blocking} are not sufficient or not efficient. We propose, then, to make offloading robust by exploiting multi-link communications for spatial error-correcting coding. Throughout Section~\ref{sec:coding}, we first define the \emph{asymmetric block-erasure channel} that models our scenario and then we analyze some of the main properties that characterize the channel and its suitable error-correcting codes. Namely, we generalize the Singleton bound to this context (Theorem~\ref{thm:singleton_bound}) and we find bounds for the outage probability (Theorem~\ref{thm:Pout_bounds}). Although we leave the explicit design of codes for the asymmetric block-erasure channel for future work, we conclude the section with a discussion on whether it is convenient \emph{to code or not to code}. We claim that under certain conditions, the use of optimal codes for multi-link offloading over the asymmetric block-erasure channel can either improve the outage probability (and hence the word error probability) during offloading or allow to considerably decrease the transmit power at the UE's side, for a given targeted outage performance.

\section{Computation offloading in 5G networks with Multi-access Edge Computing }
\label{sec:offloading}
As recalled in the introduction, computation offloading is a key application in MEC to minimize the energy consumption of mobile handsets enhancing their batteries' lifetime, or to enable heavy computation tasks in resource-poor devices such as sensors. Offloading a certain application requires an overall delay $D_{\rm off}$ defined as follows:
\begin{equation}
\label{eq:offloading_delay}
D_{\rm off}=D_{\rm tx}+D_{\rm exe}+D_{\rm rx},
\end{equation} 
where $D_{\rm tx}$ is the time needed to transfer the application and the input data from the UE to the MEH, $D_{\rm exe}$ is the time needed to run the application in the MEH, and $D_{\rm rx}$ is the time needed to get the result of the computation back. If we denote by $n_b$ the number of bits to upload from the UE to the MEH, by $w$ the number of CPU cycles necessary to run the application, and by $f_S$ the computational capacity of the MEH, expressed in CPU cycles/s, then we have $D_{\rm tx}=\frac{n_b}{B R}$ and $D_{\rm exe}=\frac{w}{f_S}$, where $R$ denotes the rate of the link in bit/s/Hz and $B$ is the fixed available bandwidth. In this article, we investigate the case of mmWave links, with the usage of beamforming techniques thanks to antenna arrays both at the transmitter and at the receiver side \cite{Kutty2016}. Given a UE (transmitter) with $n_T$ antennas and an AP (receiver) with $n_R$ available antennas, we denote by $\mathbf{H} \in \mathbb{C}^{n_R \times n_T}$ the channel matrix.
The MIMO channel capacity is given by the Shannon formula as follows:
\begin{equation}
\label{capacity_mimo}
C(\mathbf{H})=B\,\log_2{\left|\mathbf{I} + \mathbf{H}\mathbf{Q}\mathbf{H}^H\mathbf{R}^{-1}\right|} \quad \text{[bit/s]},
\end{equation}
where $|\cdot|$ denotes the determinant of $(\cdot)$, $\mathbf{Q}$ represents the covariance matrix of the transmitted symbols, and $\mathbf{R}$ is the noise covariance matrix. If we assume a white Gaussian uncorrelated noise, we can substitute $\mathbf{R}$ with a scaled identity matrix $\sigma_n^2\mathbf{I}$. After some classical and straightforward algebraic manipulations, the capacity can be written as \cite{Bar_MIMO}: 
\begin{equation*}
C(\mathbf{H}) = B\sum_{i=1}^{r}\log_2\left(1+\frac{h_i\gamma_i}{\sigma_n^2}\right),
\end{equation*}
where $r$ is the rank of $\mathbf{H}^H\mathbf{H}$, $h_i$ is the $i$-th eigenvalue of $\mathbf{H}^H\mathbf{H}$ and $\gamma_i$ is the $i$-th eigenvalue of $\mathbf{Q}$.  
Due to the large antenna gains and the narrow beams that characterize mmWave communications, for the sake of this paper we will only consider a single line-of-sight path between the UE and the AP, without secondary paths; 
this implies that $\mathbf{H}$ has rank $1$. Therefore, we can write the channel capacity as
\begin{equation}
\label{capacity_rango1}
C= C(\mathbf{H}) = B\log_2(1+ap),
\end{equation}
where $a = \frac{h_1}{\sigma_n^2}$ is the channel response incorporating the beamforming gain, divided by the noise power, and $p = \gamma_1$ is the transmit power.

One of the main goals of this paper is to minimize the UE's energy consumption during computation offloading under the latency constraint $D_{\off} \leq L$, for some constant $L>0$ measured in seconds. In general, recalling \eqref{eq:offloading_delay} and calling $R$ the uplink communication bit rate, the problem can be formulated as follows:
\begin{mini}
{}{p}
{\label{Escalar}}
{p_{\min} = }
\addConstraint{D_{\off} = \frac{n_b T_b}{R} + \frac{w}{f_S} + D_{\rm rx} \leq L}
\addConstraint{0 \leq p \leq P_T,}
\end{mini}
where $P_T$ is the maximum allowed transmit power and $T_b=1/B$.
Notice that the latency constraint is equivalent to guarantee a minimum bit rate
\begin{equation}
\label{eq:R_min}
  R \geq \frac{n_b T_b}{L - \frac{w}{f_S} - D_{\rm rx}} =: R_{\min}.
\end{equation}
Clearly, the transmission power and the communication rate are related. Given (\ref{capacity_rango1}), let us suppose that the coherence time of the channel is long enough to imply that $a$ does not change in the time interval needed by the UE to perform offloading; let us also suppose that the UE can choose an optimal modulation and coding scheme that achieves the maximum transmission rate. In this case, the minimum transmit power $p_{\min}$ that guarantees the latency constraint is simply
\begin{equation}
  \label{eq:p_min}
  p_{\min} = \frac{2^{R_{\min}} -1}{a}.
\end{equation}
If $p_{\min} \leq P_T$, offloading can be performed; otherwise, \eqref{Escalar} simply has no feasible solutions.

\section{Multi-link communications to reduce energy consumption}
\label{sec:multi_link}
The first novelty of our paper is the introduction of a new degree of freedom to the scenario described in the previous section, so that the transmit power at the UE's level can be further decreased and reduced with respect to~\eqref{eq:p_min}. To this end, we investigate the convenience of exploiting simultaneous multi-link communications between the UE and the edge cloud and we provide sufficient and necessary conditions under which using an optimal number $N^*$ (possibly greater than $1$) of simultaneous links guarantees the minimization of the UE's transmit power. 

\begin{figure}[t]
\centering
\includegraphics[width=\columnwidth]{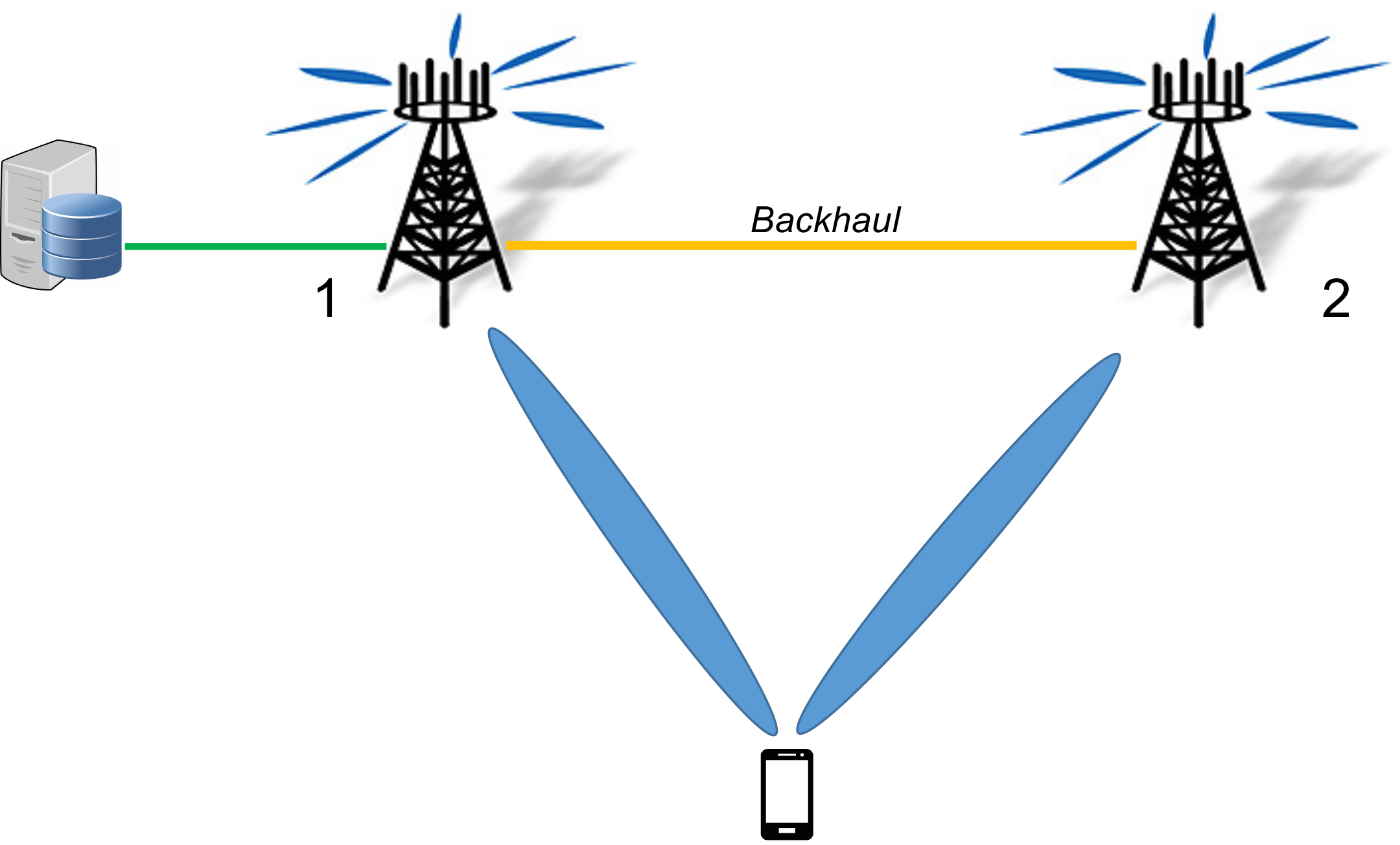}
\caption{Two-link communication between a UE and the edge cloud.}
\label{2links}
\end{figure}

From now on, when we speak of multi-link communications, we mean that the UE can send different information to different APs via different mmWave beams and over all the available UE-AP links simultaneously. This requires the use of digital beamforming. As before, we will only consider a single line-of-sight path between the UE and each of the APs, neglecting potential secondary paths. We also assume that all the APs can communicate among themselves with negligible latency through an ideal high-capacity backhaul. In this way, one AP endowed with a MEH can collect all the information sent by the UE within a negligible delay. This scenario is consistent with cloud-RAN architecture, where the APs are simple RRHs and the information is processed in the cloud. This is represented in Fig.~\ref{2links} for the case of two links.
A detailed description of multi-beam technologies in mmWave communications with fixed subarray and full multi-beam antennas is provided in \cite{Hong17}. Although therein the perspective lies on the AP's side, we consider in this work the case of UEs capable of exploiting these (or equivalent) technologies. The aim of this section is to show that in this scenario, multi-link communications can be convenient, because they allow to reduce the overall transmit power with respect to the single-link case under the same latency constraint, thus broadening the set of feasible solutions of~\eqref{Escalar}. 

Let us start by considering the double-link case: let us suppose that the user transmits $n_1$ bits over the first link with rate $R_1$ and $n_2$ bits over the second link with rate $R_2$, under the condition $n_1+n_2=n_b$. Let us suppose that the capacities of the two channels are $C_i = B \log_2(1+a_ip_i),\ i=1,2$, where the coefficients $a_i$ that identify the channel are obtained as $a$ in~\eqref{capacity_rango1}. We suppose that the UE can measure or estimate the coefficients $a_i$'s before offloading its application and that they remain constant during the whole uplink transmission phase. Moreover, without loss of generality, we can consider the first channel to be better than the second: $a_1 \geq a_2$. 

The first problem we address is how to split the total number of bits $n_b$ across the two channels and how to allocate the transmit powers so that $p_1+p_2 < p_{\min}^{(1)}$, where $p_i=\frac{2^{R_i}-1}{a_i},\ i=1,2$, is the transmit power over the $i$-th link and $p_{\min}^{(1)} = \frac{2^{R_{\min}}-1}{a_1}$ is the minimum transmit power needed to respect the latency constraint when offloading is performed only on one (the best) channel, exactly as in~\eqref{eq:p_min}. Notice that we are assuming that $a_1$ is the same both in the single-link and in the double-link case; in other words, we are assuming that the quality of the first channel (the link between the UE and the first AP) does not change in presence or absence of communications over the second channel. 

When we simultaneously send $n_1$ bits over one link and $n_2 = n_b - n_1$ bits over the second, the associated uplink transmission delay is:
\begin{equation*}
  D_{\rm tx} = \max \left( \frac{n_1 T_b}{R_1}, \frac{n_2 T_b}{R_2} \right) = \max \left( \frac{n_1 T_b}{R_1}, \frac{(n_b - n_1)T_b}{R_2} \right). 
\end{equation*}
So, the latency constraint $D_{\off} \leq L$ can be written as
\begin{align*}
  \max \left( \frac{n_1 T_b}{R_1}, \frac{(n_b - n_1)T_b}{R_2} \right) & + \frac{w}{f_S} + D_{\rm rx} \\
  & \leq L = \frac{n_b T_b}{R_{\min}} + \frac{w}{f_S} + D_{\rm rx},
\end{align*}
with $R_{\min}$ as in \eqref{eq:R_min}.
The constraint is satisfied if and only if 
\begin{equation}
  \label{eq:latency}
  \max \left( \frac{n_1}{R_1}, \frac{n_b - n_1}{R_2} \right) \leq \frac{n_b}{R_{\min}}.
\end{equation}
Hence, we are looking for $n_1, n_2, R_1, R_2$ that solve the following minimization problem:
\begin{mini}
{n_1, n_2, R_1, R_2}{p_1 + p_2}
{\label{eq:min_prob_2}}{}
\addConstraint{\max \left( \frac{n_1}{R_1}, \frac{n_b - n_1}{R_2} \right) \leq \frac{n_b}{R_{\min}}}{}{}
\addConstraint{n_1 + n_2 = n_b,}{}{n_1, n_2 \in \mathbb{N}}
\addConstraint{R_1, R_2 \geq 0}{}{}
\addConstraint{p_1 + p_2 < p_{\min}^{(1)}.}{}{}
\end{mini}
First of all, notice that the optimal solution of the problem is such that $\frac{n_1}{R_1}=\frac{n_b - n_1}{R_2}.$ Indeed, for every feasible solution such that $\frac{n_1}{R_1} < \frac{n_b - n_1}{R_2} = D_{\rm tx}$, we can always reduce $R_1$ to a lower value $R_1' < R_1$ such that $\frac{n_1}{R_1'} = D_{\rm tx}$. This decreases the transmit power on the first channel and leads to a (strictly) better solution of the problem. The same argument can be applied to a solution with $D_{\rm tx} = \frac{n_1}{R_1} > \frac{n_b - n_1}{R_2}$. Therefore, we must have $\frac{n_1}{R_1} = \frac{n_b - n_1}{R_2}$, which easily leads to
\begin{equation}
\label{eq:n1_n2}
n_1=\frac{n_b R_1}{R_1+R_2},\ \ \ n_2=\frac{n_b R_2}{R_1+R_2},\ \ \ D_{\rm tx}=\frac{n_b}{R_1+R_2}.
\end{equation}
Notice that, to keep the notation light, we are assuming that $n_1$ and $n_2$ in~\eqref{eq:n1_n2} are always integers, avoiding the use of upper and lower integer parts. At this point, we can rewrite the latency constraint as
\begin{equation*}
  \frac{n_b}{R_1 + R_2} \leq \frac{n_b}{R_{\min}} \text{\ \ \ or, equivalently,\ \ \ } R_1 + R_2 \geq R_{\min}.
\end{equation*}
Since the transmit power is an increasing function of $R_1$ and $R_2$, the optimal solution of our problem must satisfy:
\begin{equation}
\label{eq:sum_rate_R_min}
R_1+R_2=R_{\min}.
\end{equation}
The optimal solution is characterized in the following lemma, whose proof is detailed in Section~\ref{sec:proof_lemma_2_link}:
\begin{lemma}
  \label{lem:2_link}
  Consider two channels characterized by $a_1$ and $a_2$, with $a_1 \geq a_2$. 
  If
  \begin{equation}
    \label{eq:cond_alpha}
    2^{R_{\min}} > \frac{a_1}{a_2},
  \end{equation}
  the minimum transmission power is achieved by transmitting $n_1 = n_bR_1/R_{\min}$ bits over the first channel with rate $R_1$ and $n_2 = n_b - n_1$ bits over the second channel with rate $R_2 = R_{\min} - R_1$, where
  \begin{equation}
    \label{eq:R_star}
    R_1 = \frac{R_{\min}}{2} + \frac{1}{2} \log_2 \left(\frac{a_1}{a_2}\right).
  \end{equation}
  In this case, the total minimum transmit power is
  \begin{equation}
  	\label{eq:power_2}
    	p_{\min}^{(2)} = \frac{2^{\frac{R_{\min}}{2}+1}}{\sqrt{a_1a_2}} - \left( \frac{1}{a_1} + \frac{1}{a_2} \right) < p_{\min}^{(1)}.
  \end{equation}
  If instead~\eqref{eq:cond_alpha} does not hold,
     the minimum transmission power is achieved by transmitting all the $n_b$ bits over the best channel with rate $R_{\min}$.
\end{lemma}

\begin{figure}[t]
  \centering
  \resizebox {\columnwidth} {!} {
  \begin{tikzpicture}

    \datavisualization [school book axes,
      axis layer/.style={}, 
      x axis = {ticks=none},
      y axis = {ticks=none},
      visualize as smooth line/.list={bound,threshold},
      style sheet=strong colors,
      legend entry options/default label in legend path/.style=straight label in legend line,
      bound={label in legend={text={$a_2 = a_1$}}},
      threshold={label in legend={text={$a_2 = a_12^{-R_{\min}}$}}},
      data/format=function]
    
    data [set=threshold]{
      var x : interval [0:5];
      func y = \value x*0.5;
    }
    
    data [set=bound]{
      var x : interval [0:4];
      func y = \value x;
    }; 
    
    \draw[->] ({atan(0.5) + 1}:4.5 ) arc [start angle={atan(0.5) + 1}, end angle=44, radius=4.5]
    node[near end, right, scale=0.6] {$R_{\min}\rightarrow 0$};
    \draw[->] ({atan(0.5) - 1}:4.5 ) arc [start angle={atan(0.5) - 1}, end angle=1, radius=4.5]
    node[near end, left, scale=0.6] {$R_{\min}\rightarrow +\infty$};

    \node at (2.5,-0.5) {$a_1$};
    \node at (-0.5,2.5) {$a_2$};

    \begin{scope}[on background layer]
      \filldraw [gray!10] (0,0) -- (4,4) -- (0,4) -- (0,0);
      \filldraw [black!10!green!30] (0,0) -- (5,2.5) -- (5,4) -- (4,4) -- (0,0);
      \filldraw [red!10] (0,0) -- (5,0) -- (5,2.5) -- (0,0);
    \end{scope}

  \end{tikzpicture}
  }
  \caption{\label{fig:qualitative_2link}The ranges of $a_1$ and $a_2$ corresponding to Lemma~\ref{lem:2_link}.}
\end{figure}
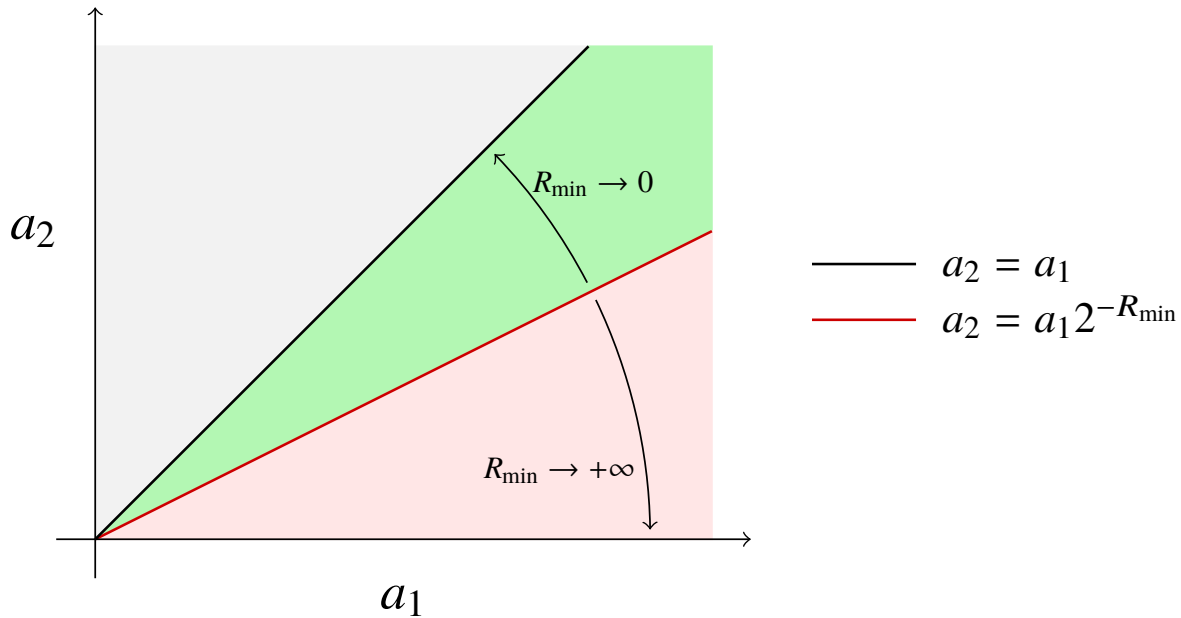
\begin{figure}[t]
\centering
\includegraphics[width=\columnwidth]{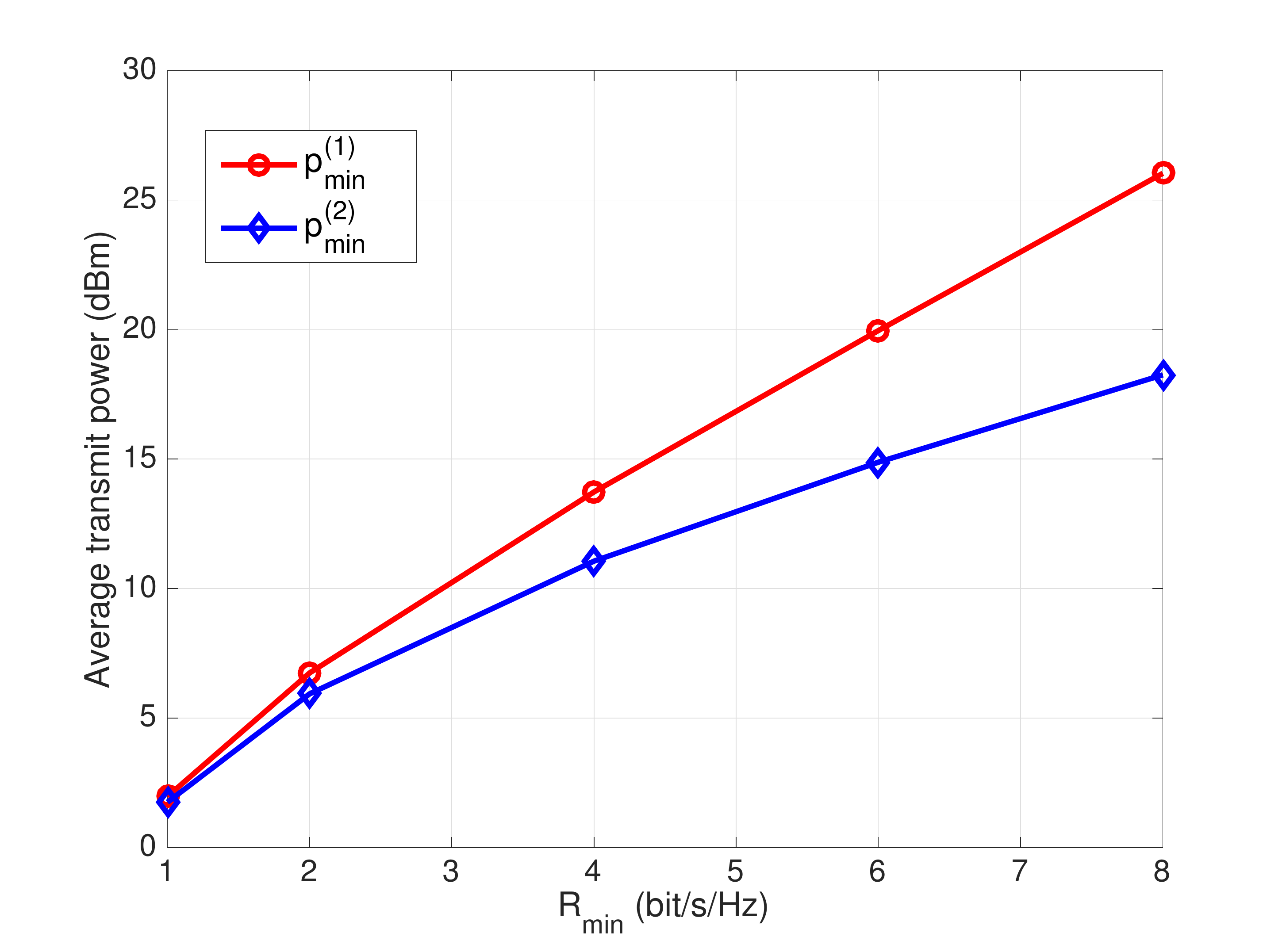}
\caption{Average minimum achievable transmit power as a function of $R_{\min}$ for the one-link and the two-link cases.}
\label{p_min_vs_p_double}
\end{figure}

Fig.~\ref{fig:qualitative_2link} depicts the region of the couples $(a_1,a_2)$ that allow power minimization through the exploitation of the double link: if $(a_1,a_2)$ lies in the green area, the double link is power-wise advantageous; if it lies in the red area, it is more convenient to use only the first channel. Notice that the tighter is the latency constraint, the higher becomes the minimum rate $R_{\min}$ necessary to guarantee it. As a consequence, the ``convenience region'' for double-link communications enlarges: somehow, we are forced to ``spread'' our information transmission over two channels, because ``squeezing'' it over only one channel with a tight latency constraint is too expensive in terms of power. 

In Fig.~\ref{p_min_vs_p_double}, we show a numerical example of how the gain in using two links increases as a function of the minimum required spectral efficiency. 
The results are averaged over randomly drawn APs' positions, uniformly distributed in a square of area $10000\ \text{m}^2$, centered at the UE's location, considering a path loss given by the Friis equation~\cite{Friis1946}. In particular, we consider that
\begin{equation}
  \label{eq:a_i}
  a_i = \frac{b}{\sigma_n^2 d_i^{\alpha}},
\end{equation}
where $d_i$ is the distance between the mobile user and the $i$-th AP (taken in ascending order: $d_1 \leq d_2$), $\alpha$ is the path loss exponent, and $b$ depends on the physical conditions of our transmission system. Friis equation in free space takes $\alpha=2$ and $b=G_RG_T\left(\frac{\lambda_w}{4\pi}\right)^2$, where $\lambda_w$ is the signal wavelength and $G_R$ and $G_T$ are the antenna gains at the receiver and transmitter side, respectively; in this case,
\begin{equation}
a_i=G_RG_T\left(\frac{\lambda_w}{4\pi \sigma_n d_i}\right)^2.  
\end{equation}
Since the APs are randomly distributed, the distances $d_i$ and hence the $a_i's$ are random variables. The random distribution of the $a_i$'s directly depends on the distribution of distances of the APs around the mobile users. For Fig.~\ref{p_min_vs_p_double}, we chose $G_R = n_R = 128$, $G_T = n_T = 32$, $\lambda_w =5$ mm, and $\sigma_n = -82,96$ dBm.

\begin{figure}[t]
\centering
\includegraphics[width=\columnwidth]{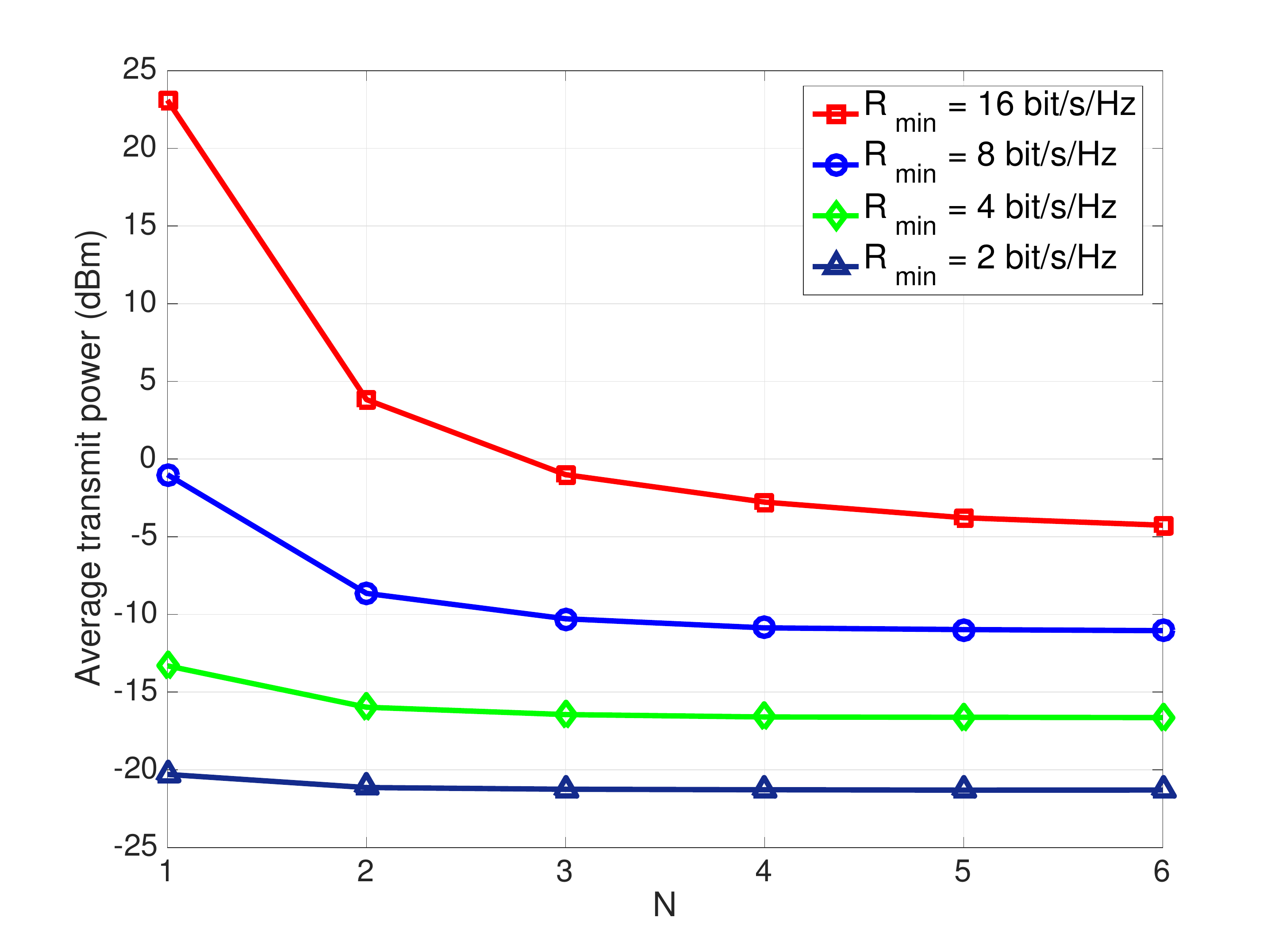}
\caption{Average transmit power for scenarios with different numbers of APs and different $R_{\min}$.
\label{pot_diff_N}}
\end{figure}

Lemma \ref{lem:2_link} can be generalized to a scenario with $N \geq 2$ available links. With Theorem~\ref{thm:N_link} and Corollary~\ref{cor:N_star}, we provide a full and explicit solution to the problem of minimizing the transmit power for computation offloading via multi-link communications. The proof of the theorem is available in Section~\ref{sec:proof_thm_N_link}.
\begin{theorem}
  \label{thm:N_link}
  Given $N$ line-of-sight links between a UE and $N$ MEC APs with capacity $C_i = B\log_2(1+a_ip_i)$ and $a_1 \geq a_2 \geq \cdots \geq a_N$, the total transmit power is minimized by simultaneously communicating over all the $N$ links (and not strictly less) if and only if
  \begin{equation}
    \label{eq:condition_N}
    2^{R_{\min}} 
    > \frac{\prod_{i=1}^{N-1}a_i}{a_N^{N-1}}.
  \end{equation}
  In this case, the (power-wise) optimal information transmission rate over the $i$-th channel is
  \begin{equation}
    \label{eq:rates}
    R_i = \frac{1}{N} \left(R_{\min} + \log_2 \left( \frac{a_i^{N-1}}{\prod_{j\neq i}a_j} \right) \right);
  \end{equation}
  $\sum_{i=1}^NR_i = R_{\min}$ and the number of bits sent over the $i$-th channel is
  \begin{equation}
    \label{eq:n_i}
    n_i = \frac{n_b R_i}{R_{\min}},
  \end{equation}
  	which guarantees that $\sum_{i=1}^Nn_i = n_b$; finally, the total minimum transmission power is
  \begin{equation}
  \label{eq:total_power}
  p_{\min}^{(N)} = \sum_{i=1}^N p_i = N \left(\frac{2^{R_{\min}}}{\prod_{i=1}^N a_i}\right)^{\frac1N} - \sum_{i=1}^N\frac{1}{a_i}.
  \end{equation}
\end{theorem}
\begin{corollary}
\label{cor:N_star}
  Given $N$ line-of-sight links between a UE and $N$ MEC APs with capacity $C_i = B\log_2(1+a_ip_i)$ and $a_1 \geq a_2 \geq \cdots \geq a_N$, the trasmit power of time-constrained computation offloading is minimized by multi-link communication over the best $N^* \leq N$ links, where $N^*$ is the only integer such that
\begin{equation}
\label{eq:cond_opt}
  \frac{a_1a_2\cdots a_{N^*-1}}{a_{N^*}^{N^*-1}} < 2^{R_{\min}} \leq \frac{a_1a_2\cdots a_{N^*}}{a_{N^*+1}^{N^*}}
\end{equation}
(imposing by convention that $a_1a_2\cdots a_{N}/a_{N+1}^{N}=+\infty$).
\end{corollary}
\begin{IEEEproof}
The proof is an immediate consequence of Theorem~\ref{thm:N_link} and the fact that
  \begin{equation}
  \label{eq:growing_coeff}
  \frac{a_1}{a_2} \leq \frac{a_1a_2}{a_3^2} \leq \cdots \leq \frac{a_1a_2\cdots a_{N-2}}{a_{N-1}^{N-2}} \leq \frac{a_1a_2\cdots a_{N-1}}{a_{N}^{N-1}}.
\end{equation}
Notice that~\eqref{eq:cond_opt} is such that~\eqref{eq:condition_N} holds for $N^*$, but not for $N^*+1$. 
\end{IEEEproof}
Fig.~\ref{pot_diff_N} shows the evolution of the average minimum achievable transmit power~\eqref{eq:total_power} as a function of the number $N$ of available APs around the user, uniformly distributed in a square of size 200 m. The results are averaged over $10000$ independent channel realizations and are plotted for different values of $R_{\min}$. The $a_i$'s are derived from the distances between the UE and the APs as in Fig.~\ref{p_min_vs_p_double}. 
According to Corollary~\ref{cor:N_star}, the user selects only the best $N^* \leq N$ channels.
Notice that, as the number $N$ of available channels increases, the average tranmsit power gain decreases, until becoming negligible when $N$ is in the order of $3$ or $4$. 


\subsection{Proof of Lemma~\ref{lem:2_link}}
\label{sec:proof_lemma_2_link}
Using the double link is convenient if and only if the sum of the transmission powers with the information split over both links is less than the transmission power needed to send all the information over the best link (the conditions imposed by~\eqref{eq:n1_n2} and~\eqref{eq:sum_rate_R_min} imply that the only truly free variable in our problem is $R_1$): 
\begin{align}
  & p_1+p_2= \frac{2^{R_1}-1}{a_1} + \frac{2^{R_2}-1}{a_2} < \frac{2^{R_{\min}}-1}{a_1} = p_{\min}^{(1)} \nonumber \\
  & \Longleftrightarrow\ \ \frac{2^{R_1}-1}{a_1} + \frac{2^{R_{\min}-R_1}-1}{a_2} < \frac{2^{R_{\min}}-1}{a_1} \nonumber \\
  & \Longleftrightarrow\ \ \frac{2^{R_1} \left(1 - 2^{R_{\min}-R_1} \right)}{a_1} < \frac{\left(1 - 2^{R_{\min}-R_1} \right)}{a_2} \nonumber \\
  \label{eq:condition}
  & \Longleftrightarrow\ \ 2^{R_1} > \frac{a_1}{a_2},
\end{align}
provided that $R_1 < R_{\min}$.

  Suppose that \eqref{eq:cond_alpha} does not hold. Since $R_1 \leq R_{\min}$, then~\eqref{eq:condition} does not hold either, $p_1 + p_2 \geq p_{\min}^{(1)}$, and it is more convenient to communicate only over the best channel. Therefore,~\eqref{eq:cond_alpha} is a necessary condition for the usefulness of exploiting also the second channel.
  
  Now, let us study the total power as a function of $R_1$, when we transmit over both channels:
  \begin{equation*}
    p_1 + p_2 = p(R_1) = \frac{2^{R_1}-1}{a_1} + \frac{2^{R_{\min}-R_1}-1}{a_2}.
  \end{equation*}
  Its first derivative is
  \begin{equation*}
    \frac{d p}{d R_1}(R_1) = \log_e2 \cdot \left( \frac{2^{R_1}}{a_1} - \frac{2^{R_{\min}-R_1}}{a_2}\right)
  \end{equation*}
  and, in particular,
  \begin{equation*}
    \frac{d p}{d R_1}(R_1) \geq 0\ \ \Longleftrightarrow\ \ R_1 \geq \frac{R_{\min}}{2} + \frac{1}{2} \log_2 \left(\frac{a_1}{a_2}\right) =: R_1^*.
  \end{equation*}
  If we suppose that~\eqref{eq:cond_alpha} holds true, we have
  \begin{equation*}
    R_{\min} > R_1^* > \log_2 \left(\frac{a_1}{a_2}\right).
  \end{equation*}
  The study of the derivative tells that it suffices to choose $R_1 = R_1^*$ to minimize $p(R_1)$ and the resulting power is strictly less than $p_{\min}^{(1)}$ (notice that $R_1= R_1^*$ satisfies~\eqref{eq:condition} by the previous inequality). Therefore,~\eqref{eq:cond_alpha} is also a sufficient condition for the usefulness of using both channels simultaneously.
When~\eqref{eq:cond_alpha} and~\eqref{eq:R_star} hold, the total transmit power is
\begin{align*}
  p_{\min}^{(2)} & = p_1 \left( \frac{R_{\min}}{2} + \frac{1}{2}\log_2 \left(\frac{a_1}{a_2}\right) \right) \\
  & \ \ \ \ \ \ \ \ \ \ + p_2 \left( \frac{R_{\min}}{2} - \frac{1}{2} \log_2 \left(\frac{a_1}{a_2}\right) \right) \\
	& = \frac{2^{\frac{R_{\min}}{2}+1}}{\sqrt{a_1a_2}} - \left( \frac{1}{a_1} + \frac{1}{a_2} \right) < p_{\min}^{(1)}.
\end{align*}

\subsection{Proof of Theorem~\ref{thm:N_link}}
\label{sec:proof_thm_N_link}

The same argument on the uplink transmission delay used to derive~\eqref{eq:n1_n2} can be applied here to 
state that $\frac{n_i}{R_i} = \frac{n_j}{R_j}$ for every $i,j \in \{1,\ldots,N\}$ such that $R_i,R_j \neq 0$. Therefore, for every fixed $i \in \{1,\ldots,N\}$, we can write $n_b = \sum_{j=1}^{N} n_j = \sum_{j=1}^{N} \frac{n_i R_j}{R_i} = \frac{n_i}{R_i}\sum_{j=1}^{N} R_j$, from which we derive that $n_i = n_b R_i \left(\sum_{j=1}^{N} R_j\right)^{-1}$. As in~\eqref{eq:sum_rate_R_min}, we can easily find that $R_1 + R_2 + \cdots + R_{N} = R_{\min}$. Therefore, \eqref{eq:n_i} is proved:
\begin{equation*}
n_i = \frac{n_b R_i}{R_{\min}}.
\end{equation*}

We will demonstrate the rest of the theorem by induction on the number of channels $N$. For $N=2$, the theorem coincides with Lemma~\ref{lem:2_link}, hence the base case is already established. For the induction step, we will assume that the optimal transmission scheme over $N$ channels obeys the statement of the theorem and we will analyze the case of $N+1$ channels. 

First of all, let us consider the case in which
\begin{equation}
\label{eq:2_min_small}
2^{R_{\min}} \leq \frac{a_1 \cdots a_{N-1}}{a_N^{N-1}}.
\end{equation} 
Suppose that there exist some $R_1, \ldots, R_{N+1}$ such that
\begin{equation}
\label{eq:p_N_plus_1}
p(R_1, \ldots, R_{N+1}) = \sum_{i=1}^{N+1} \frac{2^{R_i}-1}{a_i} \leq p_{\min}^{(N)},
\end{equation}
where $p(R_1, \ldots, R_{N+1})$ is the transmission power used for simultaneous offloading over all the $N+1$ channels and $p_{\min}^{(N)}$ is the minimum transmit power achievable with communication over only the best $N$ channels.
Let us define
\begin{equation}
\label{eq:rate_constraint}
\overline{R_{\min}} := R_{\min} - R_{N+1} = R_1 + R_2 + \cdots + R_{N}.
\end{equation}
The power minimization problem over $N$ links with rate constraint~\eqref{eq:rate_constraint} is solved by some $\overline{R_1}, \ldots, \overline{R_N}$ such that $\overline{R_j} = 0$ for some $j$. Indeed, since $\overline{R_{\min}} \leq R_{\min}$, by inductive hypothesis and by~\eqref{eq:2_min_small}, the minimum power is obtained by communicating over strictly less than $N$ links. Let us call this minimum
\begin{equation*}
\overline{p_{\min}^{(N)}} = \sum_{i=1}^{N} \frac{2^{\overline{R_i}}-1}{a_i}.
\end{equation*}
In particular, we have
\begin{equation*}
\overline{p_{\min}^{(N)}} \leq \sum_{i=1}^{N} \frac{2^{R_i}-1}{a_i},
\end{equation*}
where the $R_i$ are taken as in~\eqref{eq:p_N_plus_1}. Now, consider the quantity
\begin{align*}
p^{(N)} & = p\left(\overline{R_1}, \ldots, \overline{R_{j-1}},R_{N+1},\overline{R_{j+1}}, \ldots, \overline{R_N} \right) \\
& = \sum_{i=1}^{j-1} \frac{2^{\overline{R_i}}-1}{a_i} + \frac{2^{R_{N+1}}-1}{a_j} + \sum_{i=j+1}^{N} \frac{2^{\overline{R_i}}-1}{a_i} \\
& = \overline{p_{\min}^{(N)}} + \frac{2^{R_{N+1}}-1}{a_j},
\end{align*}
where the rate $R_{N+1}$ is the same as in~\eqref{eq:p_N_plus_1} and it replaces $\overline{R_j} = 0$. Since $a_j \geq a_{N+1}$, we have:
\begin{align*}
p^{(N)} & = \overline{p_{\min}^{(N)}} + \frac{2^{R_{N+1}}-1}{a_j} \leq \overline{p_{\min}^{(N)}} + \frac{2^{R_{N+1}}-1}{a_{N+1}} \\
& \leq \sum_{i=1}^{N} \frac{2^{R_i}-1}{a_i} + \frac{2^{R_{N+1}}-1}{a_{N+1}} \\
& = p(R_1, \ldots, R_{N+1}) \leq p_{\min}^{(N)}.
\end{align*}
Since $\left(\overline{R_1}, \ldots, \overline{R_{j-1}},R_{N+1},\overline{R_{j+1}}, \ldots, \overline{R_N} \right)$ is a feasible solution to our power minimization problem over $N$ links with constraint $\sum_{i=1}^N R_i = R_{\min}$, we must have
\begin{equation*}
p^{(N)} = p(R_1, \ldots, R_{N+1}) = p_{\min}^{(N)}.
\end{equation*}
This means, in particular, that the $(N+1)$-th link is not useful for decreasing the total transmit power: at best, we can only get as low as $p_{\min}^{(N)}$, the minimum over $N$ links. In other words, we have just proved that a necessary condition for the convenience of exploiting in parallel all the $N+1$ channels is that $2^{R_{\min}} > a_1 \cdots a_{N-1}/a_{N}^{N-1}$. From now on, let us assume it. In this case, for every fixed $R_{N+1}$, by inductive hypothesis, 
$\overline{p_{\min}^{(N)}}$ is achieved by choosing
\begin{equation}
\begin{split}
\label{eq:optimal_Ri}
\overline{R_i} & = \overline{R_i}\left(R_{N+1}\right) \\
& = \frac{1}{N}\left(\overline{R_{\min}} + \log_2\left(\frac{a_i^{N-1}}{a_1 \cdots a_{i-1}a_{i+1}\cdots a_N}\right) \right), 
\end{split}
\end{equation}
for every $i=1, \ldots,N$.
Let us look at $p_1 + \cdots + p_{N+1} = p\left(R_{N+1}\right)$ as a function of $R_{N+1}$. Let us call $R_{N+1}^*$ the value of $R_{N+1}$ that minimizes $p\left(R_{N+1}\right)$ and $R_i^* = \overline{R_i}(R_{N+1}^*)$. By convexity of $p(R_1,\ldots, R_{N+1}) = \sum_{i=1}^{N+1} (2^{R_i}-1)/a_i$, we have:
\begin{align*}
p(R_1^*, & \ldots,R_{N+1}^*) \\
& \leq p\left(\overline{R_1}(R_{N+1}),\ldots,\overline{R_N}(R_{N+1}), R_{N+1}\right)\  \forall R_{N+1} \\
& \leq p(R_1,\ldots, R_{N+1})\ \forall R_1, \ldots, R_{N+1}.
\end{align*}
Thus, the solution of the transmit power minimization problem with $N+1$ available links is $p(R_1^*,\ldots,R_{N+1}^*)$ and it is completely characterized if we know $R_{N+1}^*$. By~\eqref{eq:total_power} and by inductive hypothesis, we have:
\begin{align*}
p\left(R_{N+1}\right) & = p\left(\overline{R_1}(R_{N+1}),\ldots,\overline{R_N}(R_{N+1})\right) + \frac{2^{R_{N+1}}-1}{a_{N+1}}\\
& = N \left(\frac{2^{R_{\min}-R_{N+1}}}{\prod_{j=1}^N a_j}\right)^{\frac1N} - \sum_{i=1}^N\frac{1}{a_i} + \frac{2^{R_{N+1}}-1}{a_{N+1}}
\end{align*}
and its derivative is
\begin{equation*}
\frac{dp}{d R_{N+1}}\left(R_{N+1}\right) = \log_e 2 \cdot \left( \frac{2^{R_{N+1}}}{a_{N+1}} - \frac{2^{\frac{R_{\min} - R_{N+1}}{N}}}{\left(\prod_{j=1}^N a_j\right)^{\frac{1}{N}}}\right).
\end{equation*}
The latter is zeroed for
\begin{equation*}
R_{N+1} = \frac{1}{N+1}\left(R_{\min} + \log_2\left(\frac{a_{N+1}^N}{\prod_{j=1}^N a_j}\right) \right).
\end{equation*}
The feasibility condition $R_{N+1} \geq 0$ implies that
\begin{equation}
\label{eq:R_N_plus_1}
\begin{split}
& R_{N+1}^* = \\
& = 
\begin{cases}
	0, & \text{ if } 2^{R_{\min}} \leq \frac{a_1 \cdots a_N}{a_{N+1}^N} \\
    \frac{1}{N+1}\left(R_{\min} + \log_2\left(\frac{a_{N+1}^N}{\prod_{j=1}^N a_j}\right) \right), & \text{ otherwise}
\end{cases}.
\end{split}
\end{equation}
Therefore, in full concordance with the statement of the theorem when $N+1$ links are available, the optimal solution exclusively involves the best $N$ channels if $2^{R_{\min}} \leq \frac{a_1 \cdots a_N}{a_{N+1}^N}$; if instead $2^{R_{\min}} > \frac{a_1 \cdots a_N}{a_{N+1}^N}$, we have that~\eqref{eq:R_N_plus_1} yields $R_{N+1}^* > 0$ as in~\eqref{eq:total_power}. Substituting $R_{N+1} = R_{N+1}^*$ into~\eqref{eq:optimal_Ri}, after a few straightforward algebraic steps, we also obtain
\begin{equation*}
R_i^* = \frac{1}{N+1}\left(R_{\min} + \log_2\left(\frac{a_i^N}{a_1 \cdots a_{i-1}a_{i+1}\cdots a_{N+1}}\right) \right).
\end{equation*}
Finally, with this choice of the rates, we can explicitly derive the expression of the total power:
\begin{align*}
p_{\min}^{(N+1)} & = \sum_{i=1}^{N+1} \frac{2^{R_i^*}-1}{a_i} \\
  & = \sum_{i=1}^{N+1} \left( 2^{\frac{R_{\min}}{N+1}} \left( \frac{a_i^{N}}{\prod_{j\neq i}a_j} \right)^{\frac{1}{N+1}} - 1 \right) \frac{1}{a_i} \\
  & = \sum_{i=1}^{N+1} \left( 2^{\frac{R_{\min}}{N+1}} \left( \frac{1}{\prod_{j=1}^{N+1} a_j} \right)^{\frac{1}{N+1}} - \frac{1}{a_i} \right) \\
  & = (N+1) \left(\frac{2^{R_{\min}}}{\prod_{j=1}^{N+1} a_j}\right)^{\frac{1}{N+1}} - \sum_{i=1}^{N+1}\frac{1}{a_i}.
\end{align*}

\section{A Probability Distribution for the Optimal Number of Links}
\label{sec:opt_link_distribution}
The aim of this section is to investigate the probability that~\eqref{eq:condition_N} is satisfied when the $a_i$'s are not deterministic. The model that we take into account considers a UE placed at the center of the Euclidean space $\R^2$, with the APs  distributed around it according to a homogeneous Poisson point process $\Phi$ with intensity $\lambda$~\cite{Chiu2013}. This is a commonly investigated scenario in the literature~\cite{Hunter2008,Weber2010,Andrews2011,Singh2011,Bai2014,ElSawy2014} and, in practice, it means that:
\begin{itemize}
  \item the probability $\prob\{\Phi(S)=k\}$ of finding $k$ points of the random point process in a bounded Borel set $S \subseteq \R^2$ is
    \begin{equation}
      \label{eq:p_p_p}
      \prob\{\Phi(S)=k\} := \frac{(\lambda \mu(S))^k}{k!}e^{-\lambda \mu(S)},
    \end{equation}
    where $\mu(S)$ indicates the standard Lebesgue measure of $S$ (its area). 
  \item If $S$ and $T$ are two disjoint Borel sets, then $\prob\{\Phi(S)=k\}$ and $\prob\{\Phi(T)=\ell\}$ are independent for every $k$ and $\ell$.
\end{itemize}
The random point process models the random geometry of our network and the channel responses $a_i$'s inherit from it a random distribution. As in~\eqref{eq:a_i}, we suppose that $a_i \propto d_i^{-\alpha}$, hence we are interested in characterizing the random distances between the UE and its sorrounding APs. We borrow from \cite{Thompson1956} the following lemma:
\begin{lemma}
  \label{lem:distance_distribution}
  Given a homogeneous Poisson point process in $\R^2$ with intensity $\lambda$, the distance $d_i$ between a fixed point of the space and its $i$-th closest point of the process is randomly distributed according to the following probability density function:
\begin{equation*}
  f_{d_i}(x) = \frac{2}{(i-1)!} (\lambda \pi)^i x^{2i-1} e^{-\lambda \pi x^2}.
\end{equation*}
The joint probability density function of $d_1 \leq d_2 \leq \ldots \leq d_N$ is
\begin{equation}
\label{eq:density_function}
  f_{d_1, \ldots, d_N}(x_1,\ldots,x_N) = (2 \lambda \pi)^N  x_1x_2 \cdots x_N e^{-\lambda \pi x^2_N}.
\end{equation}
\end{lemma}

Now, if $a_i \propto d_i^{-\alpha}$, \eqref{eq:condition_N} holds true if and only if
\begin{equation}
  \label{eq:condition_distances}
  d_N^{N-1} < d_1 d_2 \cdots d_{N-1} 2^{\frac{R_{\min}}{\alpha}}.
\end{equation}
We are interested in characterizing the probability that~\eqref{eq:condition_distances} holds true depending on the distribution of the distances $d_i$'s. From now on, we introduce the notation
\begin{equation*}
  A := 2^{\frac{R_{\min}}{\alpha}} > 1.
\end{equation*}
$\mathcal{E}_N := \{d_N^{N-1} < d_1 d_2 \cdots d_{N-1} A\}$ represents the event that~\eqref{eq:condition_distances} holds true. As in Corollary~\ref{cor:N_star}, let $N^*$ be the number of links that minimizes the transmission power in multi-link computation offloading. If $\overline{\mathcal{E}_N}$ denotes the complementary event of $\mathcal{E}_N$, then, as a consequence of Corollary~\ref{cor:N_star}, the probability that $N^* = N$ is
\begin{align*}
  \prob\{N^* = N\} & = \prob\{\mathcal{E}_N \cap \overline{\mathcal{E}_{N+1}}\} \\
  & = \prob \left \{ \frac{d_{N}^{N-1}}{d_1d_2\cdots d_{N-1}} < A \leq \frac{d_{N+1}^N}{d_1d_2\cdots d_N} \right \}.
\end{align*}

\begin{figure}
\centering
\includegraphics[width=\columnwidth]{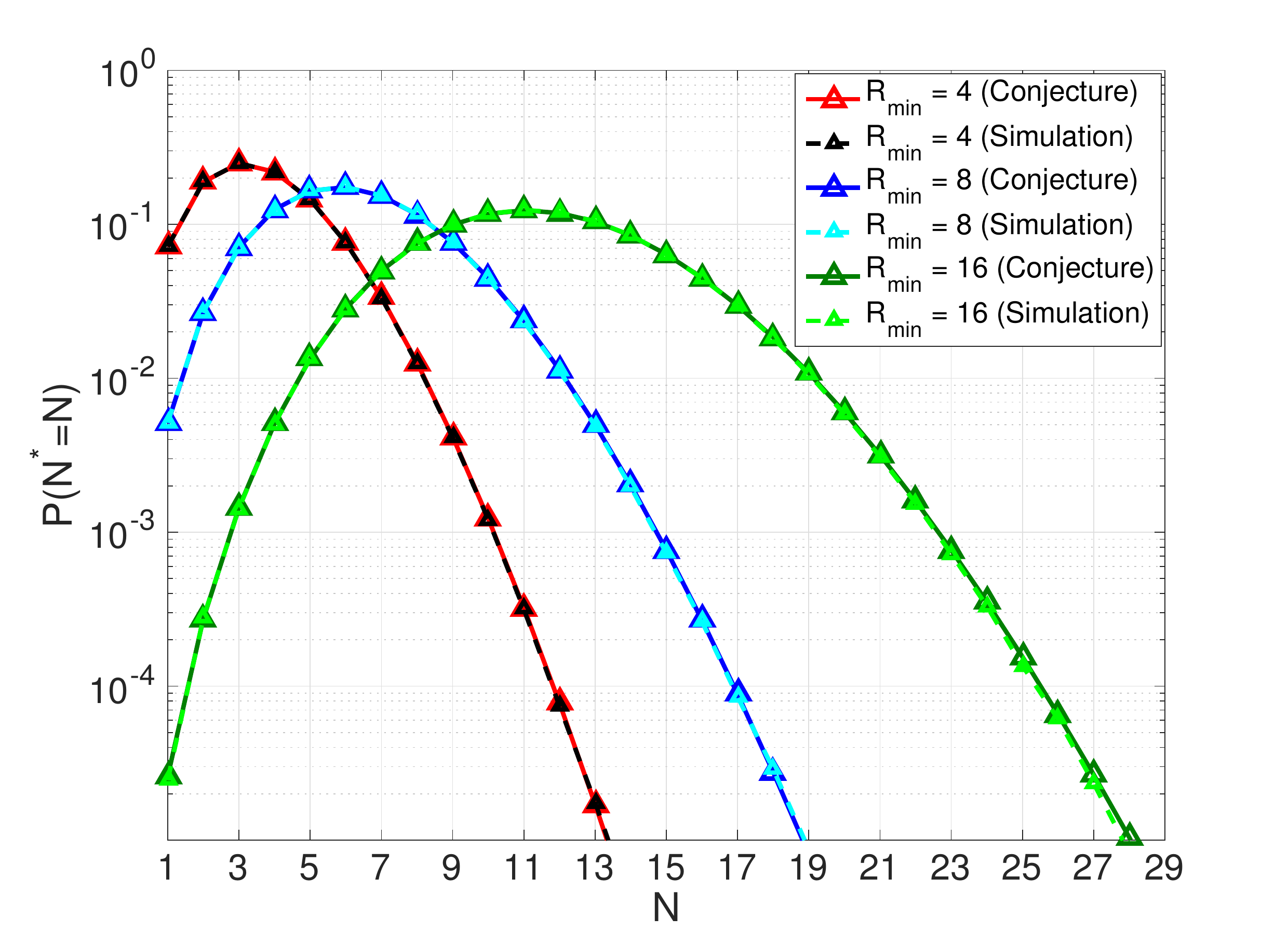}
\caption{Comparison between the formula for $\prob\{N^*=N\}$ given by Conjecture~\ref{conj:N_opt} and its value measured through numerical simulations.}
\label{fig_conj_vs_sim}
\end{figure}

\begin{figure}
\centering
\includegraphics[width=\columnwidth]{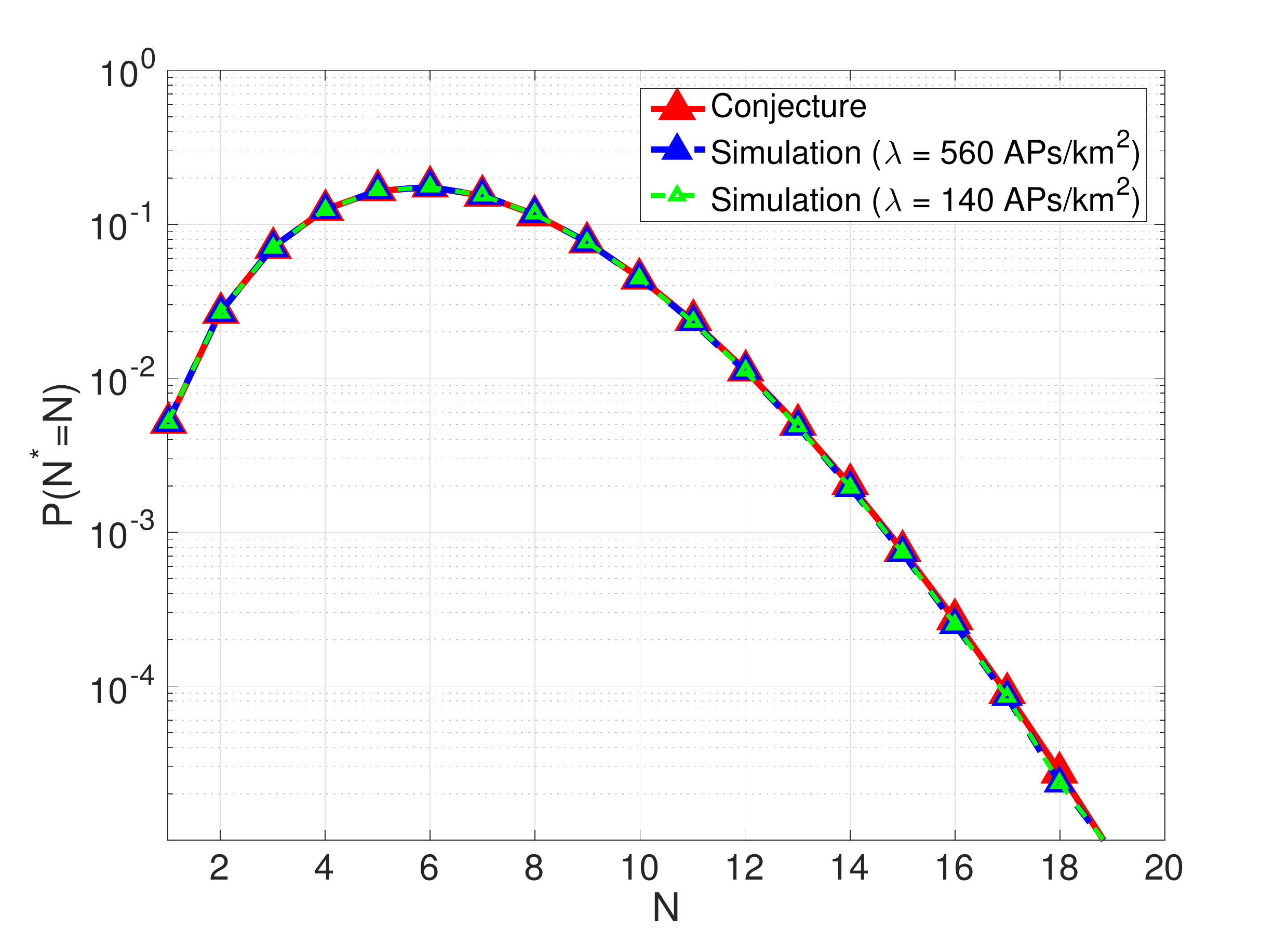}
\caption{Numerical evidence of the independence of $\prob\{N^* = N\}$ from the APs' density $\lambda$.}
\label{fig_conj_vs_sim_lambda}
\end{figure}

Since $d_1 \leq d_2 \leq \ldots \leq d_N$, the events $\mathcal{E}_N$ are included in one another. Indeed,
\begin{align}
  & d_N^{N-1} < d_1 d_2 \cdots d_{N-1} A\nonumber \\ 
  & \Rightarrow\ \  d_{N-1}^{N-1} < d_N^{N-1} < d_1 d_2 \cdots d_{N-1} A\nonumber \\
  \label{eq:dist_ineq}
  & \Rightarrow\ \ d_{N-1}^{N-2} < d_1 d_2 \cdots d_{N-2} A\ \\
  & \Rightarrow\ \ d_{N-2}^{N-2} < d_{N-1}^{N-2} < d_1 d_2 \cdots d_{N-2} A\nonumber \\
  & \ \ \ \ \ \ \ \ \ \ \vdots\ \ \ \ \ \vdots\ \ \ \ \ \vdots\ \ \ \ \ \vdots\ \ \ \ \ \vdots\ \ \ \ \ \vdots\ \ \ \ \ \nonumber \\
  & \Rightarrow\ \ d_3^2 < d_1 d_2 A\ \ \Rightarrow\ \ d_2^2 < d_3^2 < d_1 d_2 A\ \Rightarrow\ \ d_2 < d_1 A. \nonumber
\end{align}
In other words, $\mathcal{E}_N\ \Rightarrow\ \mathcal{E}_{N-1}\ \Rightarrow\ \cdots\ \Rightarrow\ \mathcal{E}_{2}$ or, equivalently, $\mathcal{E}_N \subseteq \mathcal{E}_{N-1} \subseteq \cdots \subseteq \mathcal{E}_2$. This implies that, for $N > 0$,
\begin{align*}
  & \prob\{N^* = N\} = \\
  & = \begin{cases}
    1 - \prob\{\mathcal{E}_2\}, & \text{if } N=1,\\
    \prob\{\mathcal{E}_N \cap \overline{\mathcal{E}_{N+1}}\} = \prob\{\mathcal{E}_N\} - \prob\{\mathcal{E}_{N+1}\}, & \text{otherwise,}
  \end{cases}
\end{align*}
and, for every $N\geq 2$,
\begin{equation*}
  \prob\{N^* \geq N\} = \prob\{\mathcal{E}_N\}.
\end{equation*}
Given the probability densities enstablished in Lemma~\ref{lem:distance_distribution}, we can explicitly compute $\prob\{N^* = 1\}$ and $\prob\{N^* = 2\}$:
\begin{lemma}
\label{lem:N_star_1_2}
In the random scenario described in this section, 
we have
\begin{equation*}
	\prob\{N^* = 1\} = \frac{1}{A^2} \text{\ \ \ and\ \ \ } \prob\{N^* = 2\} = \frac{\log_eA^2}{A^2}.
\end{equation*}
\end{lemma}
The proof of the previous lemma is detailed in Section~\ref{sec:proof_conjecture}. 
More generally, given Lemma~\ref{lem:distance_distribution} and \eqref{eq:dist_ineq}, it can be straightforwardly shown that, for every $N>1$, we have
\begin{align*}
  & \prob\{N^* = N\} = \prob\{\mathcal{E}_N \cap \overline{\mathcal{E}_{N+1}}\} = \\
  & = \int_{0}^{+\infty} \int_{x_1}^{x_1A} \int_{x_2}^{\sqrt{x_1x_2A}} \cdots \int_{x_{N-1}}^{\sqrt[N-1]{x_1x_2\cdots x_{N-1}A}} \cdot \\
  & \ \ \ \ \ \cdot \int_{\sqrt[N]{x_1x_2\cdots x_NA}}^{+\infty} f(x_1,x_2,\ldots,x_{N+1}) dx_{N+1}\cdots dx_2 dx_1,
\end{align*}
where $f(x_1,\ldots,x_{N+1})$ is the version with $N+1$ variables of~\eqref{eq:density_function}.
In particular, solving the most internal integral on $x_{N+1}$, we obtain:
\begin{equation}
\label{eq:big_integral}
\begin{split}
  & \prob\{N^* = N\} = \\
  & = \int_{0}^{+\infty} \int_{x_1}^{x_1A} \int_{x_2}^{\sqrt{x_1x_2A}} \cdots \int_{x_{N-2}}^{\sqrt[N-2]{x_1x_2\cdots x_{N-2}A}} \cdot \\
  & \ \ \ \ \ \cdot \int_{x_{N-1}}^{\sqrt[N-1]{x_1x_2\cdots x_{N-1}A}} (2\lambda \pi)^N x_1x_2\cdots x_N \cdot \\
  & \ \ \ \ \ \ \ \ \ \ \cdot e^{-\lambda \pi (x_1x_2\cdots x_NA)^{\frac{2}{N}}} dx_N\cdots dx_2 dx_1.
\end{split}
\end{equation}
The previous integral cannot be easily solved into a closed-form expression for $N \geq 3$. Nonetheless, we formulate the following conjecture: 
\begin{conjecture}
  \label{conj:N_opt}
  For every $N \in \mathbb{N}\smallsetminus\{0\}$, we have:
  \begin{equation}
    \label{eq:conjecture}
    \prob\{N^* = N\} = \frac{(\log_eA^2)^{N-1}}{(N-1)!A^2}.
  \end{equation}
  That is, the random variable $N^* -1$ follows a Poisson distribution with parameter $\log_eA^2 = \frac{2R_{\min}}{\alpha \log_2e}$. Therefore, $\mathbb{E}[N^*] = \var(N^*) = \frac{2R_{\min}}{\alpha \log_2e} + 1$. 
\end{conjecture}
We conducted several numerical direct evaluations of the integral in~\eqref{eq:big_integral} for several choices of $N$, $A$, and $\lambda$; the results always coincided with the conjectured solution~\eqref{eq:conjecture}. In addition, the validity of Conjecture~\ref{conj:N_opt} is strongly supported by Fig.~\ref{fig_conj_vs_sim}, which is not based on the direct calculation of~\eqref{eq:big_integral}; instead, the values of $\prob\{N^* = N\}$ are obtained by drawing $10^7$ random realizations of the APs' positions and each time calculating the corresponding $N^*$ according to Corollary~\ref{cor:N_star}, for different values of $R_{\min}$. It is interesting to notice that~\eqref{eq:conjecture} does not depend on $\lambda$, the intensity of the Poisson point process modeling the geometry of the APs around the user. This independence is validated by the numerical simulation results depicted in Fig.~\ref{fig_conj_vs_sim_lambda}, obtained with $10^7$ random realizations of the APs' positions, for $R_{\min}=8$ and different densities $\lambda$.


In our scenario, by Conjecture~\ref{conj:N_opt}, the probability that the power-wise optimal number of links $N^*$ does not exceed $N$ is
\begin{equation*}
  \prob\{N^* < N+1\} = \sum_{i=1}^N \prob\{N^* = i\} = \sum_{i=1}^N \frac{(\log_eA^2)^{i-1}}{(i-1)!A^2}.
\end{equation*}
Therefore, it is possible to estimate (with certitude as close to $1$ as wanted) the number $N$ of links that a UE needs to ``see'' around itself to guarantee that the transmission power can be minimized, i.e.~that $N^* \leq N$. This straightforwardly yields a targeted base station density for our deployment. 
More precisely, given $\varepsilon > 0$, let us define
\begin{equation}
  \label{eq:Me}
  M_{\varepsilon} = \min \{N : \prob\{N^* \leq N\} \geq 1-\varepsilon\}.
\end{equation}
A few values of $M_{\varepsilon}$ are shown in Table~\ref{tab:M_eps}.
\begin{table}
\renewcommand{\arraystretch}{1.3}
  \caption{$M_{0.1}$ and $M_{0.01}$ for some choices of $R_{\min}$ and given $\alpha = 2$.}
  \label{tab:M_eps}
  \centering
    \bgroup
    \begin{tabular}{|c||c|c|c|c|c|c|}
      \hline
      $R_{\min}$ (bit/s/Hz)  & $0.5$ & $1$ & $2$ & $4$ & $8$  & $16$ \\
      \hline
      {\bf $M_{0.1}$}  & $2$   & $3$ & $4$ & $6$ & $10$  & $16$ \\      
      \hline
      {\bf $M_{0.01}$} & $3$   & $4$ & $6$ & $8$ & $12$ & $20$ \\
      \hline
    \end{tabular}
    \egroup
\end{table} 
$M_{\varepsilon}$ has an operational meaning: when designing an AP deployment, we may want to ensure as much as possible that users never ``see'' around them less than $M_{\varepsilon}$ APs. Indeed, with this choice, we would guarantee that with probability at least $1-\varepsilon$, the UE will be able to select the multi-link communication strategy over $N^*$ links that minimizes its transmit power for computation offloading.
These considerations allow us to estimate a minimum desired or ``targeted'' deployment density of APs: let us suppose that the UE is such that, in absence of blocking, efficient communication is possible with any AP within a range of $r$ meters; any AP further than $r$ meters is too far away and the user will not even try to communicate with it. Let us call $B(r)$ a circle of radius $r$ centered at the user. In the notation of~\eqref{eq:p_p_p}, $\Phi(B(r))$ is the (random) number of points of our Poisson process that lie inside $B(r)$. We can write:
\begin{align*}
  \prob\{\Phi(B(r)) \geq M_{\varepsilon}\} & = 1 - \prob\{\Phi(B(r)) \leq M_{\varepsilon}-1\} \\
  & = 1 - e^{-\lambda \pi r^2} \left( \sum_{i=0}^{M_{\varepsilon}-1} \frac{(\lambda \pi r^2)^{i}}{i!} \right).
\end{align*}
Now, given $\delta >0$, let us define 
\begin{align}
  \lambda_{\varepsilon,\delta} & = \inf \{\lambda > 0 : \prob\{\Phi(B(r)) \geq M_{\varepsilon}\} \geq 1-\delta\} \nonumber \\
  \label{eq:cond_lambda}
  & = \inf \left \{ \lambda > 0 : \sum_{i=0}^{M_{\varepsilon}-1} \frac{(\lambda \pi r^2)^{i}}{i!} \leq \delta e^{\lambda \pi r^2} \right \}.
\end{align}
Notice that $\lambda_{\varepsilon,\delta}$ is well-defined, because for every $\delta$, the exponential function on the right side of the inequality in~\eqref{eq:cond_lambda} always ends up growing faster in $\lambda$ than the polynomial on the left side. Under these premises, if the Poisson point process describing the AP deployment has intensity $\lambda_{\varepsilon,\delta}$, then 
\begin{align*}
  & \prob\{N^* \leq \Phi(B(r))\} \\
  & \geq \prob\{N^* \leq \Phi(B(r))\ |\ \Phi(B(r)) \geq M_{\varepsilon}\}\prob\{\Phi(B(r)) \geq M_{\varepsilon}\}\\
  & \geq \prob\{N^* \leq \Phi(B(r))\ |\ \Phi(B(r)) \geq M_{\varepsilon}\}(1 - \delta) \\
  & \geq \prob\{N^* \leq M_{\varepsilon}\ |\ \Phi(B(r)) \geq M_{\varepsilon}\}(1 - \delta) \\
  & = \prob\{N^* \leq M_{\varepsilon}\}(1 - \delta) \geq (1 - \varepsilon)(1 - \delta).
\end{align*}
In other words, the probability that at least $N^*$ APs are ``visible'' by the UE within its communication range $r$ is at least $(1-\varepsilon)(1-\delta)$. Thus, $\lambda_{\varepsilon,\delta}$ represents the minimum required AP density to make the UE's chances to optimize its transmission power as high as a given threshold, specified by $\varepsilon$ and $\delta$. 
As an example, the values of $\lfloor \lambda_{0.1,0.1} \rfloor$ are reported in Table~\ref{tab:density} for a communication range of $r=100$ m. If, for practical applications, the values of $M_{\varepsilon}$ or $\lambda_{\varepsilon,\delta}$ derived via~\eqref{eq:Me} and~\eqref{eq:cond_lambda} turn out to be too high, we can decrease them by recalling the comment we made at the end of Section~\ref{sec:multi_link}: practically, we may prefer to exploit $N < N^*$ links for offloading, because the power gain brought by the use of additional links is negligible. In such a case, we can reformulate~\eqref{eq:Me} to obtain a smaller $M_{\varepsilon}$ and, consequently, less dense deployments.


\begin{table}
\renewcommand{\arraystretch}{1.3}
  \caption{$M_{0.1}$ and $\lfloor \lambda_{0.1,0.1} \rfloor$ for some choices of $R_{\min}$ and given $\alpha = 2, r=100 \text{ m}$.}
  \label{tab:density}
  \centering
    \bgroup
    \begin{tabular}{|c||c|c|}
      \hline
          $R_{\min}$ (bit/s/Hz) & $M_{0.1}$ & $\lfloor \lambda_{0.1,0.1} \rfloor$ in APs/km$^2$ \\
          \hline \hline  
          $0.5$ & $2$  & $123$ \\
          $1$   & $3$  & $169$ \\
          $2$   & $4$  & $212$ \\
          $4$   & $6$  & $295$ \\
          $8$   & $10$ & $452$ \\
          $16$  & $16$ & $677$ \\
          \hline
    \end{tabular}
    \egroup
\end{table} 

\subsection{Proof of Lemma~\ref{lem:N_star_1_2}}
\label{sec:proof_conjecture}
Let us start with the following:
\begin{align}
  \prob\{\mathcal{E}_2\} & = \prob\{d_2 < d_1A\} = \prob\{N^* \geq 2\} \nonumber\\
  & = \int_{0}^{+\infty} \int_{x_1}^{x_1A} f_{d_1,d_2}(x_1,x_2) dx_2dx_1 \nonumber\\
  \label{eq:pE2}
  & = \int_{0}^{+\infty} \int_{x_1}^{x_1A} (2 \lambda \pi)^2  x_1x_2 e^{-\lambda \pi x_2^2} dx_2dx_1 \\
  & = \int_{0}^{+\infty} 2 \lambda \pi x_1 \left [ - e^{-\lambda \pi x_2^2} \right]_{x_1}^{x_1A} dx_1 \nonumber\\
  \label{eq:first_term}
  & = \int_{0}^{+\infty} 2 \lambda \pi x_1 e^{-\lambda \pi x_1^2} dx_1 + \\
  \label{eq:second_term}
  & \ \ \ \ \ - \int_{0}^{+\infty} 2 \lambda \pi x_1 e^{-\lambda \pi A^2 x_1^2} dx_1. 
\end{align}
By Lemma~\ref{lem:distance_distribution}, \eqref{eq:first_term} is equal to $\prob\{0 < d_1 < +\infty\} = 1$, whereas \eqref{eq:second_term} equals $A^{-2}\prob\{0 < d_1 < +\infty\} = A^{-2}$ (this can be easily seen by substituting $y = x_1A$ in the integral). Therefore,
\begin{equation*}
  \prob\{\mathcal{E}_2\} = 1 - \frac{1}{A^2} = 1 - 2^{-\frac{2R_{\min}}{\alpha}}
\end{equation*}
and
\begin{equation*}
  \prob\{N^* = 1\} = 1 - \prob\{\mathcal{E}_2\} = \frac{1}{A^2} = 2^{-\frac{2R_{\min}}{\alpha}}.
\end{equation*}

Concerning the event $\mathcal{E}_3$, we have:
\begin{align*}
  \prob\{\mathcal{E}_3\} & = \prob\{d_3^2 < d_1d_2A\} = \prob\{N^* \geq 3\}\\
  & = \int_{0}^{+\infty} \int_{x_1}^{+\infty} \prob\{d_3 < \sqrt{d_1d_2A}\ |\ d_1=x_1, d_2=x_2\} \cdot \\
  & \ \ \ \ \ \ \ \ \ \ \ \ \ \ \ \ \ \ \ \ \ \ \cdot f_{d_1,d_2}(x_1,x_2) dx_2dx_1.
\end{align*}
Now, we know by definition that $d_2^2 < d_3^2$ and we are investigating the probability that $d_3^2 < d_1d_2A$. Are the two inequalities consistent? Only if $d_2^2 < d_1d_2A$ or, equivalently, $d_2 < d_1A$ (which is the condition defining $\mathcal{E}_2$). Therefore, 
\begin{equation*}
  \prob\{d_3 < \sqrt{d_1d_2A}\ |\ d_1=x_1, d_2=x_2\} = 0, \text{ when } x_2 \geq x_1A
\end{equation*}
and the integral becomes:
\begin{align}
   & \prob\{\mathcal{E}_3\} = \int_{0}^{+\infty} \int_{x_1}^{x_1A} \prob\{d_3 < \sqrt{d_1d_2A}\ |\ d_1=x_1, d_2=x_2\} \cdot \nonumber \\
   & \ \ \ \ \ \ \ \ \ \ \ \ \ \ \ \ \ \ \ \ \ \ \ \ \ \ \ \ \ \ \ \ \cdot f_{d_1,d_2}(x_1,x_2) dx_2dx_1 \nonumber \\
   & = \int_{0}^{+\infty} \int_{x_1}^{x_1A} \int_{x_2}^{\sqrt{x_1x_2A}} f_{d_1,d_2,d_3}(x_1,x_2,x_3) dx_3dx_2dx_1 \nonumber \\
  & = \int_{0}^{+\infty} \int_{x_1}^{x_1A} (2 \lambda \pi)^2x_1x_2 \cdot \nonumber \\
  & \ \ \ \ \ \ \ \ \ \ \ \ \ \ \ \ \ \ \ \cdot \int_{x_2}^{\sqrt{x_1x_2A}} 2 \lambda \pi x_3 e^{-\lambda \pi x_3^2} dx_3dx_2dx_1\nonumber \\
  & = \int_{0}^{+\infty} \int_{x_1}^{x_1A} (2 \lambda \pi)^2x_1x_2 \left [ - e^{-\lambda \pi x_3^2} \right]_{x_2}^{\sqrt{x_1x_2A}}dx_2dx_1\nonumber \\
   \label{eq:last_integral_1}
  & = \int_{0}^{+\infty} \int_{x_1}^{x_1A} (2 \lambda \pi)^2  x_1x_2 e^{-\lambda \pi x_2^2} dx_2dx_1 + \\
  & \ \ \ \ - \int_{0}^{+\infty} \int_{x_1}^{x_1A} (2 \lambda \pi)^2  x_1x_2 e^{-\lambda \pi x_1x_2A} dx_2dx_1.
  \label{eq:last_integral_2}
\end{align}
Notice that~\eqref{eq:last_integral_1} equals~\eqref{eq:pE2}, hence it is equal to $\prob\{\mathcal{E}_2\} = 1 - A^{-2}$. Moreover,~\eqref{eq:last_integral_2} can be explicitly computed:
\begin{align}
  & \int_{0}^{+\infty} \int_{x_1}^{x_1A} (2 \lambda \pi)^2  x_1x_2 e^{-\lambda \pi x_1x_2A} dx_2dx_1 = \nonumber \\
  & = \int_{0}^{+\infty} \frac{4 e^{-\lambda \pi x_1^2A}(\lambda \pi x_1^2A + 1)}{x_1A^2} dx_1 + \nonumber \\
  & \ \ \ \ \ - \int_{0}^{+\infty} \frac{4 e^{-\lambda \pi x_1^2A^2}(\lambda \pi x_1^2A^2 + 1)}{x_1A^2} dx_1 \nonumber \\
  & = \lim_{\varepsilon \rightarrow 0} \Bigg( \int_{\varepsilon}^{+\infty} \frac{4 e^{-\lambda \pi x_1^2A}(\lambda \pi x_1^2A + 1)}{x_1A^2} dx_1 + \nonumber \\
  & \ \ \ \ \ \ \ \ \ - \int_{\varepsilon}^{+\infty} \frac{4 e^{-\lambda \pi x_1^2A^2}(\lambda \pi x_1^2A^2 + 1)}{x_1A^2} dx_1 \Bigg) \nonumber\\
  & = \lim_{\varepsilon \rightarrow 0} \Bigg( \int_{\varepsilon}^{+\infty} \frac{4 e^{-\lambda \pi x_1^2A}(\lambda \pi x_1^2A + 1)}{x_1A^2} dx_1 + \nonumber \\
  \label{eq:variable_change}
  & \ \ \ \ \ \ \ \ \ - \int_{\varepsilon \sqrt{A}}^{+\infty} \frac{4 e^{-\lambda \pi y^2A}(\lambda \pi y^2A + 1)}{yA^2} dy \Bigg) \\
  & = \frac{4}{A^2} \lim_{\varepsilon \rightarrow 0} \underbrace{\int_{\varepsilon}^{\varepsilon \sqrt{A}} \frac{e^{-\lambda \pi x_1^2A}(\lambda \pi x_1^2A + 1)}{x_1} dx_1}_{\text{\normalsize $I_{\varepsilon}$}}, \nonumber
\end{align}
where~\eqref{eq:variable_change} is obtained substituting $y = x_1\sqrt{A}$ in the second integral. Now, notice that the function $e^{-\lambda \pi x_1^2A}(\lambda \pi x_1^2A + 1)$ is strictly decreasing when $x_1 > 0$, therefore:
\begin{align*}
  e^{-\lambda \pi \varepsilon^2A^2}(\lambda \pi \varepsilon^2A^2 & + 1) \int_{\varepsilon}^{\varepsilon \sqrt{A}}\frac{1}{x_1}dx_1 \leq I_{\varepsilon} \\
  & \leq e^{-\lambda \pi \varepsilon^2A}(\lambda \pi \varepsilon^2A + 1) \int_{\varepsilon}^{\varepsilon \sqrt{A}}\frac{1}{x_1}dx_1.
\end{align*}
Since $\int_{\varepsilon}^{\varepsilon \sqrt{A}}\frac{1}{x_1}dx_1 = \frac{\log_e{A}}{2}$, we obtain:
\begin{equation*}
  \lim_{\varepsilon \rightarrow 0} I_{\varepsilon} = \frac{\log_eA}{2}
\end{equation*}
and, from~\eqref{eq:last_integral_1} and~\eqref{eq:last_integral_2}, we conclude that 
\begin{equation*}
  \prob\{\mathcal{E}_3\} = \prob\{\mathcal{E}_2\} - \frac{2\log_eA}{A^2} = 1 - \frac{1 + 2\log_eA}{A^2}.
\end{equation*}
This allows us to compute
\begin{equation*}
  \prob\{N^* = 2\} = \prob\{\mathcal{E}_2\} - \prob\{\mathcal{E}_3\} = \frac{\log_eA^2}{A^2}.
\end{equation*}

\section{Multi-link Communications and Resource Overprovisioning to Counteract Blocking}
\label{sec:multilink_with_blocking}
As anticipated in Section~\ref{sec:introduction}, one of the major drawbacks of mmWave communications is that they are prone to blocking events, due to human body, obstacles, and high interference in case of beam alignment with other users~\cite{Abouelseoud2013,Singh2011}. In this and the next section, we propose two ways to combine the power-optimization advantages of multi-link offloading with the compensation of blocking effects. In general, we can differentiate between long-term blocking events, whose duration is almost as long as the uplink transmission time of the offloading procedure (or even more, up to a few seconds~\cite{MacCartney2017}), and short-term blocking events that instead last much less. The latter can be caused, for example, by a bicycle or a car rapidly crossing the communication path between the UE and an AP. When this happens, a mmWave channel suffers from a high attenuation that temporarily decreases the achievable rate from~\eqref{capacity_rango1} to almost $0$. Substantially, a mmWave link assumes an ``on/off'' behaviour depending on the absence or presence of a physical obstacle interrupting the line-of-sight communication path. Thus, brief blocking events essentially make communication intermittent, especially in the scenario without multi-paths that we are considering in this paper. To counteract this effect, we present an approach that was first introduced in~\cite{BarbarossaCeci2017} and~\cite{Barbarossa2017}. This idea is based on overprovisioning of radio resources to guarantee an actual average information transmission rate that takes into account blocking probabilities and compensates possible information losses. 
Since the power consumption required in this case is higher than in absence of blocking, it is useful to combine overprovisioning with the multi-link communication techniques presented in the previous sections.


Several models exist that quantify the blocking probability of mmWave signals~\cite{Thornburg2016,Gapeyenko2016,Andrews2017,Qi2017,Ghatak2017}. Motivated by the stochastic AP deployment considered in Section~\ref{sec:opt_link_distribution}, let us recall the model proposed in~\cite{Bai2014}: in the bidimensional space, obstacles are assumed to be rectangles with random length $X$, width $W$, and centers randomly distributed according to a Poisson point process with density $\mu$. Then, the probability that the line-of-sight communication path between the UE and an AP at distance $d$ is \emph{not} obstructed is~\cite[Corollary 1.1]{Bai2014}:
\begin{equation}
 \label{eq:p_on}
 P_{\on}(\mu,d) = \exp(-\beta d - q),
\end{equation}
where $\beta = 2 \mu \pi^{-1} (\mathbb{E}[W] + \mathbb{E}[X])$ and $q = \mu \mathbb{E}[W] \mathbb{E}[X]$. In the rest of the section, for simplicity, we restrict our presentation to statistically independent blocking events and to the double-link scenario of Fig.~\ref{2links}. The extension to a general $N$-link scenario 
can be derived along the same lines as the sequel. An investigation of this problem in case of statistically dependent blocking events is given in \cite{BarbarossaCeci2017}. 

Now, the probability that the $i$-th channel is blocked is $P_i=1-P_{\on}(\mu,d_i)$, for $i=1,2$, and, assuming statistical independence of blocking events, during offloading the UE can experiment the following four different situations:
\begin{itemize}
\item the first link is open and the second is blocked, happening with probability $\mathbb{P}_1^{(1)} := (1-P_1)P_2$;
\item the first link is blocked and the second is open, happening with probability $\mathbb{P}_2^{(1)} := P_1(1-P_2)$;
\item both links are open, happening with probability $\mathbb{P}_1^{(2)} := \mathbb{P}_2^{(2)} := (1-P_1)(1-P_2)$;
\item both links are blocked, happening with probability $P_1P_2$.
\end{itemize}
We suppose that, during the offloading procedure, whenever a link exploited by the UE gets blocked, the UE stops transmitting over that channel (and, if possible, it adjusts the transmit power over the other channel). Moreover, whenever a formerly blocked link opens up, the UE can allocate part or all of its remaining data transmission over that link. Finally, when both links are open at the same time, the UE can choose whether to exploit them simultaneously for double-link communications or just use the best among them. Of course, if both links are blocked, the UE must interrupt its transmission and wait until at least one of the two opens up. In the rest of the section we will call $p_i^{(1)}$ the transmit power allocated for communication over the $i$-th channel when the other is blocked (for $i=1,2$) and $p_i^{(2)}$ the power allocated over the $i$-th channel when both links are open; the latter is the case of simultaneous double-link transmission, in which the total power equals $p_1^{(2)} + p_2^{(2)}$. Obviously, the transmission power is null when both links are blocked. 

As pointed out in \cite{Barbarossa2017}, we can interpret the loss of received information experienced by the AP and due to blocking events as a decrease in the average rate. Indeed, over each channel, the achievable rate drops to zero during the blocking events and can go back to a positive value only when the obstacle causing the blocking moves away. Therefore, short-term blocking can be seen as a reduction of the average information transmission rate over the whole uplink transmission time interval. When at least one channel is open, the allocated transmit power needs to be tuned to compensate the impossibility to communicate during the instants when all channels are blocked. With this aim, assuming ergodicity, we can formulate a power optimization problem analogous to~\eqref{Escalar} and~\eqref{eq:min_prob_2}, as the minimization of the average transmit power under the constraint of guaranteeing a minimum average uplink transmission rate $\bar{R}$:

\begin{mini}
{\mathbf{p}}{\sum_{j=1}^2\sum_{i=1}^2 \mathbb{P}_i^{(j)} p_i^{(j)}}
{\label{opt_blocking}}{}
\addConstraint{\bar{R}:=\sum_{j=1}^2\sum_{i=1}^2\mathbb{P}_i^{(j)} \log_2(1+a_i p_i^{(j)})\geq R_{\rm min}}{}{}
\addConstraint{p_i^{(j)}\ge 0,}{}{i=1,2, \quad j=1,2}
\addConstraint{p_i^{(1)}\le P_T,}{}{i=1,2}
\addConstraint{\sum_{i=1}^2 p_i^{(2)}\leq P_T,}{}{}
\end{mini}
where $\mathbf{p} = \left( p_1^{(1)},p_1^{(2)},p_2^{(1)},p_2^{(2)} \right)$ is the vector of powers and $a_i$ is, as in the previous sections, the $i$-th channel response. The Lagrangian associated to this constrained problem is
\begin{align*}
{\mathcal L} &= \sum_{j=1}^2\sum_{i=1}^2 \mathbb{P}_i^{(j)} p_i^{(j)} - \gamma \left( \bar{R} - R_{\min} \right) -\sum_{j=1}^2\sum_{i=1}^2 \alpha_i^{(j)} p_i^{(j)}\\
&+\sum_{i=1}^2 \nu_i \left( p_i^{(1)}-P_T \right)+\nu_3 \left( p_1^{(2)}+p_2^{(2)}-P_T \right),
\end{align*}
where $\gamma$, $\alpha_i^{(j)}, i=1,2,\, j=1,2$, and $\nu_i, i=1,2,3$, are the Lagrange multiplier associated to the constraints of~\eqref{opt_blocking}.
The Karush-Kuhn-Tucker conditions can be expressed as follows:
\begin{align*}
&a) \quad\nabla_{p_1^{(1)}}{\cal L}=\mathbb P_1^{(1)} - \frac{\gamma \mathbb P_1^{(1)}\,a_1}{(1+a_1 p_1^{(1)})\log_e 2} - \alpha_1^{(1)} + \nu_1=0;\\
&b) \quad\nabla_{p_2^{(1)}}{\cal L}=\mathbb P_2^{(1)} - \frac{\gamma \mathbb P_2^{(1)}\,a_2}{(1+a_2 p_2^{(1)})\log_e 2} - \alpha_2^{(1)} + \nu_2=0;\\
&c) \quad\nabla_{p_1^{(2)}}{\cal L}=\mathbb P_1^{(2)} - \frac{\gamma \mathbb P_1^{(2)}\,a_1}{(1+a_1p_1^{(2)})\log_e 2} - \alpha_1^{(2)} + \nu_3=0;\\
&d) \quad\nabla_{p_2^{(2)}}{\cal L}=\mathbb P_2^{(2)} - \frac{\gamma \mathbb P_2^{(2)}\,a_2}{(1+a_2p_2^{(2)})\log_e 2} - \alpha_2^{(2)} + \nu_3=0;\\
&e) \quad \gamma \left( \bar{R} - R_{\min} \right)=0,\quad\gamma\geq0,\quad \bar{R} \ge R_{\min};\\
&f) \quad\alpha_i^{(j)}p_i^{(j)}=0,\quad\alpha_i^{(j)}\geq0,\quad p_i^{(j)}\geq0,\,i=1,2,\;j=1,2;\\
&g) \quad\nu_i(p_i^{(1)}-P_T)=0,\quad\nu_i\geq0,\quad p_i^{(1)}\leq P_T ,\,i=1,2;\\
&h) \quad\nu_3(p_1^{(2)}+p_2^{(2)}-P_T)=0,\quad\nu_3\geq0,\quad p_1^{(2)}+p_2^{(2)}\leq P_T.
\end{align*}
From the first four conditions, we can write
\begin{align*}
    p_i^{(1)} & =\frac{\gamma\mathbb{P}_i^{(1)}}{(\mathbb{P}_i^{(1)}-\alpha_i^{(1)}+\nu_i)\log_e{2}}-\frac{1}{a_i},\;\; i=1,2, \\
    p_i^{(2)} & =\frac{\gamma\mathbb{P}_i^{(2)}}{(\mathbb{P}_i^{(2)}-\alpha_i^{(2)}+\nu_3)\log_e{2}}-\frac{1}{a_i},\;\; i=1,2.
\end{align*}
Interestingly, whenever the solution of the problem is such that none of the $p_i^{(j)}$'s is null and the total power does not reach the maximum power budget $P_T$, then $\gamma$ can be expressed in closed form. Indeed, when the constraints of~\eqref{opt_blocking} related to the $p_i^{(j)}$'s hold with inequality sign, we set $\nu_i=0, i=1,2,3$, and $\alpha_i^{(j)}=0$, so that
\begin{equation}
\label{closedpow}
p_i^{(j)}=\frac{\gamma}{\log_e 2}-\frac{1}{a_i},\quad i=1,2,\;j=1,2.
\end{equation}
In this case, the power allocated on one open link is always the same independently on the state of the other link. 
Note that $\gamma$ is necessarily different from zero, otherwise the powers would not respect the condition to be non-negative, and it can be determined by imposing that the condition on the transmission rate in~\eqref{opt_blocking} holds with the equality sign ($\bar{R} = R_{\min}$). By replacing \eqref{closedpow} therein, we have
\begin{equation}
\label{closedgamma}
\begin{split}
&\sum_{j=1}^2\sum_{i=1}^{2}\mathbb P_i^{(j)}\log_2\left(\frac{\gamma\,a_i}{\log_e{2}}\right)=\log_2\left(\frac{\gamma}{\log_e{2}}\right)\sum_{j=1}^2\sum_{i=1}^{2}\mathbb P_i^{(j)}+\\ &+\sum_{j=1}^2\sum_{i=1}^{2}\mathbb P_i^{(j)}\log_2(a_i)=R_{\min}.
\end{split}
\end{equation}
Since $\sum_{j=1}^2\sum_{i=1}^{2}\mathbb P_i^{(j)}=2-P_1-P_2$, from \eqref{closedgamma} we get a closed form for $\gamma$ as follows:
\begin{equation}
\label{gamma}
\displaystyle \gamma=\displaystyle(\log_e{2})\cdot2^ {\displaystyle  \left(\frac{R_{\min}-\sum_{i=1}^{2}(1-P_i)\log_2(a_i)}{2-P_{1}-P_{2}}\right)}.
\end{equation}
\eqref{closedpow} and \eqref{gamma} show the advantage of using multi-link communications in terms of transmit power: indeed, from (\ref{gamma}) it is obvious that the transmit powers increase dramatically if and only if both links are often blocked, that means $P_{1}$ and $P_{2}$ are close to $1$. Instead, if at least one link is rarely blocked, $P_{1}$ or $P_{2}$ is close to zero and multi-link communications help in reducing power consumption.

\begin{figure}[t]
\centering
\includegraphics[width=\columnwidth]{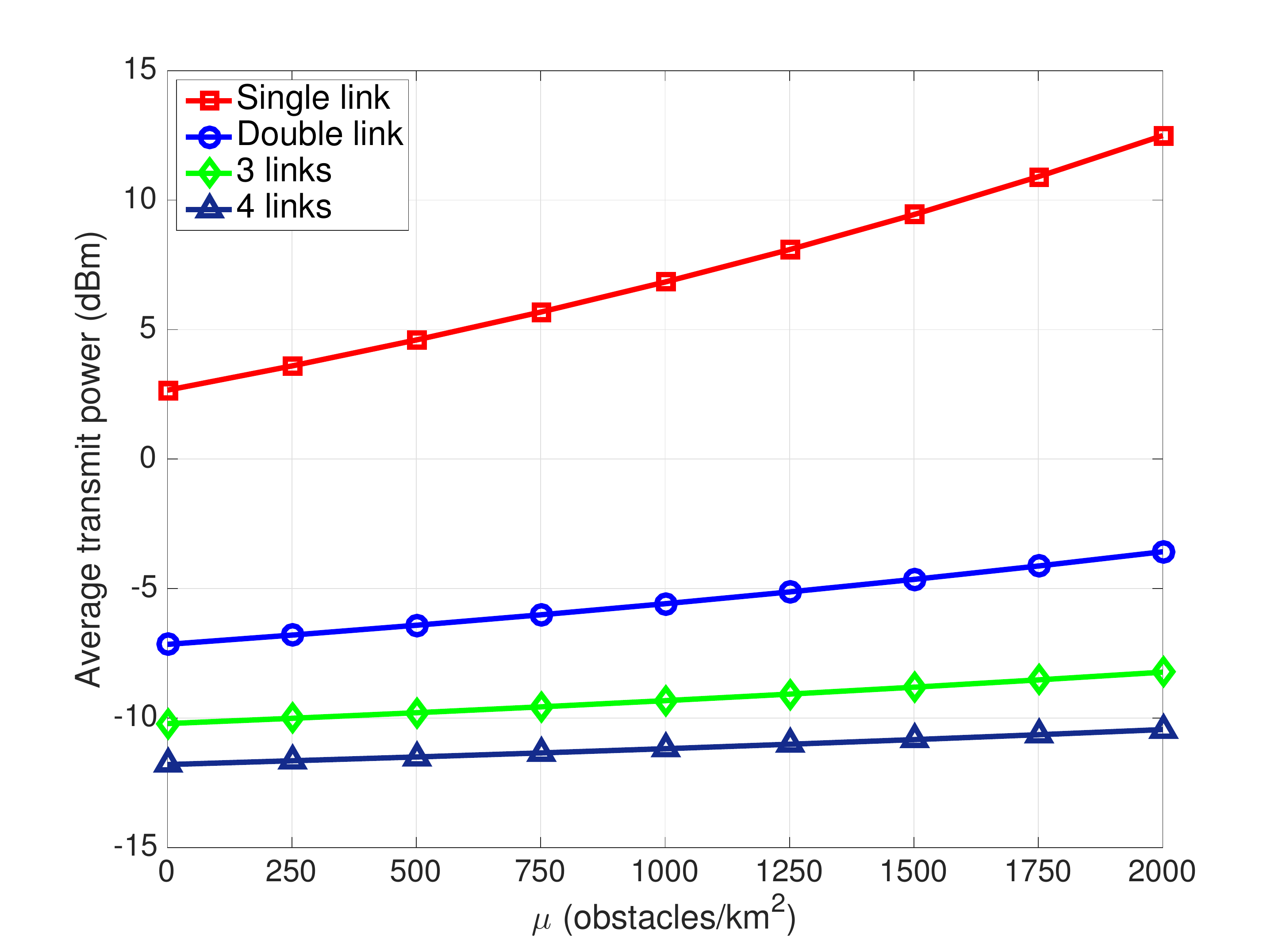}
\caption{Average transmit power as a function of the density of obstacles.}
\label{fig:ML_blocking}
\end{figure}

The formulation of the problem as it is developed in this section can be generalized to the case with more than two links. 
Both with two or more available links, the optimization problem is convex and its actual solution can be obtained with efficient numerical tools~\cite{cvxr}; as a numerical example, in Fig.~\ref{fig:ML_blocking} we show how the average transmit power depends on the number of links and on the density of obstacles in the serving area. The transmit power is averaged over random realizations of the APs' position, uniformly distributed in a square of size $150$ m. The average sizes of the obstacles are $\mathbb{E}[W]=\mathbb{E}[X]=2\ \text{m}$ and the blocking probabilities obey~\eqref{eq:p_on}. From the figure, we can notice how the highest gain is achieved in passing from $1$ to $2$ links. Moreover, the slope of the curves shows that exploiting more than one link diminishes the sensitivity of the system to the increase of the density of obstacles, i.e. the system suffers less from the blocking probability on each link.

\section{Block-Erasure-Correcting Codes for Robust Multi-link Communications}
\label{sec:coding}
In the previous section, we proposed a method to contrast short-term blocking events, which ``cut'' a mmWave communication link only for a period much shorter than the transmission time. Conversely, when blocking events last longer, overprovisioning is not effective anymore and other solutions need to be explored. This can happen when obstacles slowly cross the line-of-sight path between the UE and the AP and obstruct the link for ``long'' time intervals, even as long as a few seconds~\cite{MacCartney2017} or more. When this happens, waiting for the channel to be open again takes too much time. One solution may be to complete the offloading procedure by restarting it over other links, but this takes time and typically violates the latency constraint.
To overcome this problem, in this section we define and analyze a theoretical framework to combine error-correcting-coding techniques with multi-link mmWave communications to simultaneously perform computation offloading and contrast long-term blocking events that start after the beginning of the offloading procedure, without the need for retransmissions. 


Let us suppose to apply the multi-link communication strategy proposed in Section~\ref{sec:multi_link} over $N$ channels, transmitting $n_i$ bits over the $i$-th channel at rate $R_i$, with $\sum_{i=1}^Nn_i=n_c$ and $n_1 \geq n_2 \geq \cdots \geq n_N$. As before, the $i$-th channel is the communication link between the UE and its $i$-th closest AP, situated at distance $d_i$. Let us call $P_i$ the blocking probability of the $i$-th channel and let us assume that the distances are ordered in decreasing sense, so that $P_1 \leq P_2 \leq \cdots \leq P_N$. This is a realistic assumption, because longer line-of-sight paths have a higher chance to be blocked. For instance, the model specified in~\eqref{eq:p_on}, with $P_i = 1 - P_{\on}(\mu,d_i)$, respects this hypothesis. Consider, for simplicity, that blocking events are mutually independent on any two channels. In this case, the problem of offloading $n_c$ bits over $N$ channels without losing information is equivalent to the problem of transmitting a word of length $n_c$ bits over an \emph{asymmetric block-erasure channel}, for which the $n_c$ bits are split into $N$ blocks of length $n_i$ bits and each block has erasure probability $P_i$. Whenever one link is blocked, we suppose that all the bits of the corresponding block are lost (erased) and this happens independently from block to block. This model is our generalization of the block-erasure channel described in \cite{Fabregas2006}. We call it ``asymmetric'' because we allow 
all the $n_i$'s and the $P_i$'s to be different from each other. Our idea is to apply block-erasure-coding to multi-link communications to counteract blocking effects 
and we start by generalizing and enriching the results of~\cite{Fabregas2006}.

Formally, let $\mathcal{C} \subseteq \{0,1\}^{n_c}$ be an error-correcting code for the asymmetric block-erasure channel of rate $R_{\mathcal{C}} = \log_2|\mathcal{C}|/n_c$. Notice that in this section we are using the notation $\sum_{i=1}^Nn_i=n_c$, calling $n_c$ the total number of (coded) transmitted bits. As in the previous section, we still denote with $n_b$ the number of uncoded information bits; we also denote with $R_{\mathcal{C}}=n_b/n_c$ the coding rate. 
The codewords of $\mathcal{C}$ are written as $\mathbf{x} = (\mathbf{x}_1|\mathbf{x}_2|\cdots|\mathbf{x}_N)$, where $\mathbf{x}_i$ is the block of $n_i$ coordinates transmitted over the $i$-th link. Let us define an \emph{erasure pattern} $\mathbf{e}$ as the vector $\mathbf{e} = (e_1,e_2,\ldots,e_N) \in \{0,1\}^{N}$ such that $e_i =1$ if the $i$-th block of a codeword is erased (i.e.~if the $i$-th UE-AP link is blocked) and $e_i =0$ otherwise. Thus, $\prob\{e_i =1\} = P_i$. For a given $\mathbf{e}$, we define
\begin{equation*}
  \mathcal{C}(\mathbf{e}) = \{ \mathbf{x} \in \mathcal{C} : \text{ if } e_i=0 \text{ then } \mathbf{x}_i=0,\ \forall i=1,2,\ldots,N \}.
\end{equation*}
$\mathcal{C}(\mathbf{e})$ is the set of codewords of $\mathcal{C}$ whose non-zero blocks are only among the erased blocks identified by $\mathbf{e}$. If $\mathcal{C}$ is a linear code, then we can suppose without loss of generality that the asymmetric block-erasure channel input is the all-zero codeword $\zero = (0,0,\ldots,0)$. For every given erasure pattern $\mathbf{e}$, all the codewords of $\mathcal{C}(\mathbf{e})$ will give the same channel output as $\zero$. Assuming that a \emph{maximum likelihood} decoder does not give priority to any of the codewords of $\mathcal{C}(\mathbf{e})$, the \emph{word error probability} caused by the erasure pattern $\mathbf{e}$ is
\begin{equation*}
  P_e^{w}(\mathbf{e}) = 1 - \frac{1}{|\mathcal{C}(\mathbf{e})|}.
\end{equation*}
In particular, if $\mathcal{C}(\mathbf{e}) = \{\zero\}$ and $|\mathcal{C}(\mathbf{e})| = 1$, the decoder is capable of correctly decoding the erasure pattern $\mathbf{e}$. Therefore, for linear codes, the word error probability associated with the $P_i$'s equals
\begin{equation*}
  P_e^{w} = P_e^{w}(P_1,\ldots,P_N) := \mathbb{E}_{\mathbf{e}}[P_e^{w}(\mathbf{e})] = \mathbb{E}_{\mathbf{e}} \left[ 1 - \frac{1}{|\mathcal{C}(\mathbf{e})|} \right],
\end{equation*}
where the expected value is computed with respect to the distribution of the erasure pattern.
We give the following definition of diversity:
\begin{definition}
  The \emph{block-diversity} of a code $\mathcal{C}$ is defined as
  \begin{equation*}
    \delta = \min_{\substack{\mathbf{x},\mathbf{y} \in \mathcal{C} \\ \mathbf{x} \neq \mathbf{y}}} |\{ i \in \{1,2,\ldots,N\} : \mathbf{x}_i \neq \mathbf{y}_i\}|.
  \end{equation*}
\end{definition}
Notice that for every erasure pattern $\mathbf{e}$ such that $\delta > \sum_{i=1}^N e_i$, there will be no ML-decoding error. Therefore, we are interested in designing codes with the biggest diversity possible. It is clear that, in general, $\delta \leq N$ and we say that a code has \emph{full diversity} if $\delta = N$. An upper bound for $\delta$ is given by our generalization of the Singleton bound defined in \cite{Fabregas2006} for the case where all blocks have the same length. In our more general setup, we have:
\begin{theorem}[Singleton bound]
\label{thm:singleton_bound}
  Let $0 < R_{\mathcal{C}} \leq 1$ and let $\ell \in \{1,\ldots,N\}$ be the only integer such that
  \begin{equation}
    \label{eq:ell}
    \sum_{i=\ell+1}^N n_i < n_c R_{\mathcal{C}} \leq \sum_{i=\ell}^N n_i.
  \end{equation}
  Let us call $M = \frac{1}{N-\ell+1}\sum_{i=\ell}^Nn_i$ the average length of the last $N-\ell+1$ blocks of a codeword. Then,
  \begin{equation*}
    \delta \leq \left\lfloor 1 + N - \frac{n_c R_{\mathcal{C}}}{M} \right\rfloor =: \delta_{\singb}.
  \end{equation*}
\end{theorem}
\begin{IEEEproof}
  Let $I$ be any subset of $\{1,2,\ldots,n_c\}$ of cardinality $\sum_{i=\ell+1}^Nn_i$. Then, there exist two codewords of $\mathcal{C}$ that coincide at least on all the coordinates indexed by the elements of $I$: indeed, let us suppose by contradiction that the opposite held true; all codewords of $\mathcal{C}$ would be different on the subset of coordinates indexed by $I$. This would imply that the cardinality of $\mathcal{C}$ could not exceed the number of possible binary vectors of length $|I|$. Therefore, using~\eqref{eq:ell}, we would obtain
  \begin{equation*}
    2^{n_c R_{\mathcal{C}}} = |\mathcal{C}| \leq 2^{|I|} = 2^{\sum_{i=\ell+1}^N n_i} < 2^{n_c R_{\mathcal{C}}},
  \end{equation*}
  which is impossible. Hence, for every $I$, there exist two codewords of $\mathcal{C}$ that coincide on the coordinates indexed by all $i \in I$. Choosing $I$ as the set of the last $N-\ell$ blocks of a codeword, we deduce that there are always two codewords of $\mathcal{C}$ that coincide on those blocks and $\delta$ cannot be greater than the number of remaining blocks: $\delta \leq \ell$. Now, using again~\eqref{eq:ell} and the definition of $M$, we have:
  \begin{equation*}
    n_c R_{\mathcal{C}} \leq (N-\ell+1)M \leq (N-\delta+1)M,
  \end{equation*}
  from which it is easy to derive that $\delta \leq \delta_{\singb}$.
\end{IEEEproof}
\begin{corollary}
  If $n_c R_{\mathcal{C}} > \frac{1}{N-\ell+1}\sum_{i=\ell}^Nn_i$, then $\delta_{\singb} < N$ and the code cannot have full diversity.
\end{corollary}

Now, let us define the \emph{outage} probability as the probability that, due to blocking events, the received number of bits is less than $n_b = n_cR_{\mathcal{C}}$ (the number of information bits):
\begin{equation}
\label{eq:p_out}
  P_{\out} = \prob \left \{ \sum_{i=1}^N (1-e_i)n_i < n_c R_{\mathcal{C}} \right \}.
\end{equation}
Obviously, in case of outage, correct decoding is impossible, regardless of the goodness of the code. Hence, $P_e^{w} \geq P_{\out}$.

\begin{theorem}
  \label{thm:Pout_bounds}
  Let $0 < R_{\mathcal{C}} \leq 1$, let $\ell \in \{1,\ldots,N\}$ be the only integer such that
  \begin{equation*}
    \sum_{i=\ell+1}^N n_i < n_c R_{\mathcal{C}} \leq \sum_{i=\ell}^N n_i
  \end{equation*}
  and, analogously, let $j \in \{0,\ldots,N-1\}$ be the only integer such that
  \begin{equation*}
    \sum_{i=1}^j n_i < n_c R_{\mathcal{C}} \leq \sum_{i=1}^{j+1} n_i.
  \end{equation*}
  The outage probability is bounded as follows:
  \begin{align*}
    P_{\out} & \geq \sum_{u=0}^j \binom{N}{u} \prod_{i=1}^{N-u} P_i \prod_{i=N-u+1}^N (1-P_i), \\
    P_{\out} & \leq \sum_{u=0}^{N-\ell} \binom{N}{u} \prod_{i=1}^{u} (1-P_i) \prod_{i=u+1}^N P_i.
  \end{align*}
\end{theorem}
\begin{IEEEproof}
  Let us start from the lower bound. Whenever the channel output consists of at most $j$ out of the $N$ blocks composing the codeword, we are in outage: indeed, recalling that $n_1 \geq n_2 \geq \cdots \geq n_N$, for every $I \subseteq \{1,2,\ldots,N\}$ such that $|I| \leq j$, by definition of $j$ we have that $\sum_{i \in I}n_i \leq \sum_{i=1}^j n_i < n_c R_{\mathcal{C}}$. Therefore,
  \begin{equation*}
    \begin{split}
    P_{\out} & \geq \prob\{|\{i : e_i = 0 \}| \leq j \} \\
    & = \sum_{u=0}^j \sum_{\substack{I \subseteq \{1,2,\ldots,N\} \\ |I|=u}} \prod_{i \in \{1,2,\ldots,N\} \smallsetminus I} P_i \prod_{i \in I} (1-P_i). 
  \end{split}
  \end{equation*}
  Now, notice that $P_1 \leq P_2 \leq \cdots \leq P_N$ implies that $P_s(1-P_t) \geq P_t(1-P_s)$ for every $t\leq s$. Therefore, for every $u$, it is less probable to receive the last $u$ blocks (and not to receive the first $N-u$) than receiving any other possible set of $u$ blocks. Hence,
  \begin{align*}
    \sum_{u=0}^j & \sum_{\substack{I \subseteq \{1,2,\ldots,N\} \\ |I|=u}} \prod_{i \in \{1,2,\ldots,N\} \smallsetminus I} P_i \prod_{i \in I} (1-P_i) \\
    & \geq \sum_{u=0}^j \sum_{\substack{I \subseteq \{1,2,\ldots,N\} \\ |I|=u}} \prod_{i=1}^{N-u} P_i \prod_{i=N-u+1}^N (1-P_i) \\
    & = \sum_{u=0}^j  \binom{N}{u} \prod_{i=1}^{N-u} P_i \prod_{i=N-u+1}^N (1-P_i),
  \end{align*}
  which is the desired lower bound.

  Concerning the upper bound, the key observation is that, whenever the channel output consists of at least $N-\ell+1$ blocks, we cannot be in outage: for every $I \subseteq \{1,2,\ldots,N\}$ such that $|I| \geq N-\ell+1$, by definition of $\ell$ we have that $\sum_{i \in I}n_i \geq \sum_{i=\ell}^N n_i \geq n_c R_{\mathcal{C}}$. Consequently, $ P_{\out} \leq \prob\{|\{i : e_i = 0 \}| \leq N-\ell \}$ and the upper bound is obtained along the same lines as the lower bound, arguing that for every $u$, it is more probable to receive the first $u$ blocks (and not to receive the last $N-u$) than receiving any other possible set of $u$ blocks.
\end{IEEEproof}

\begin{figure}
\centering
\includegraphics[width=\columnwidth]{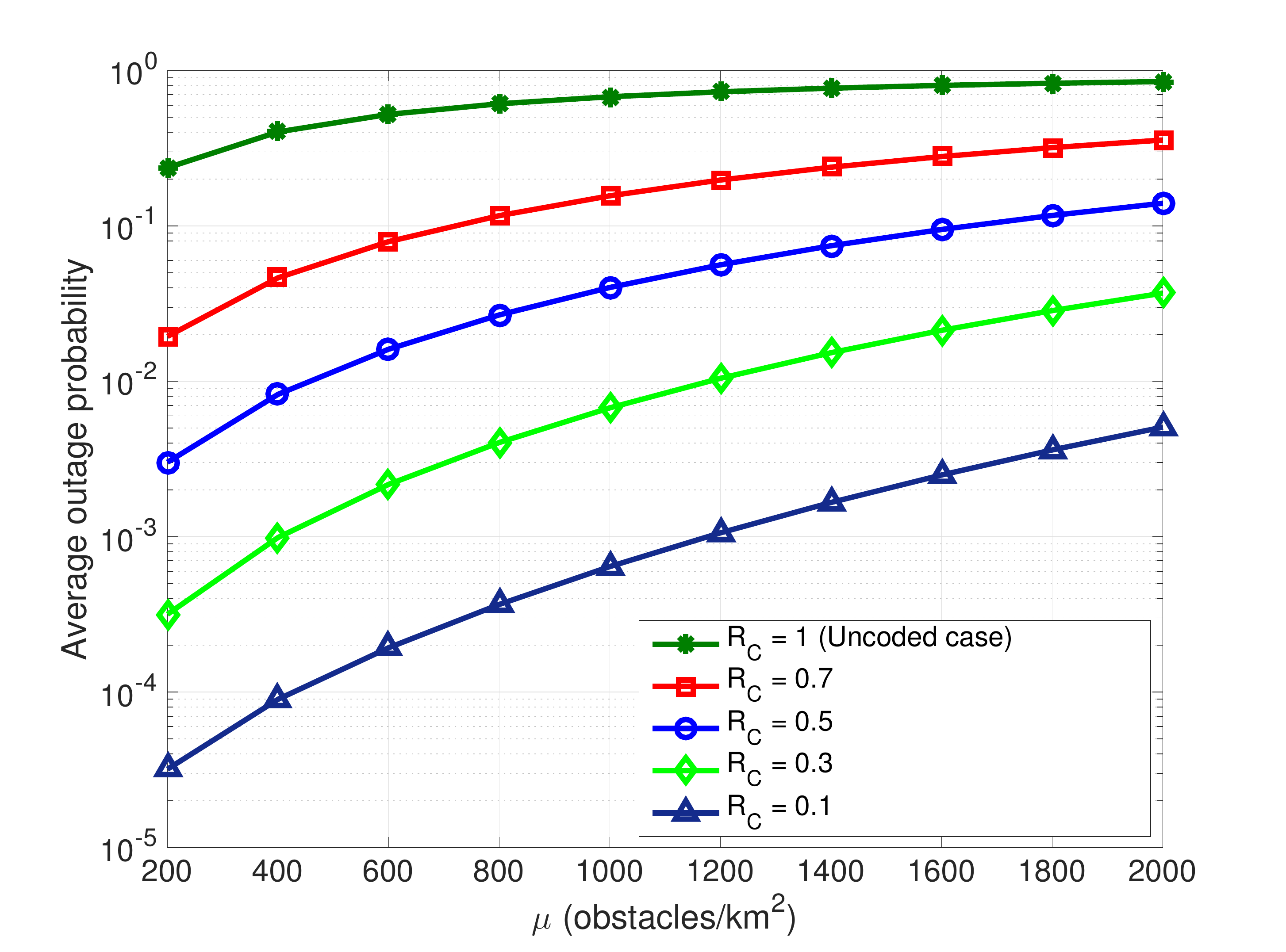}
\caption{Average outage probability as a function of the density of obstacles, for different values of $R_{\mathcal{C}}$.}
\label{fig:outage_vs_density}
\end{figure}

\begin{figure}
\centering
\includegraphics[width=\columnwidth]{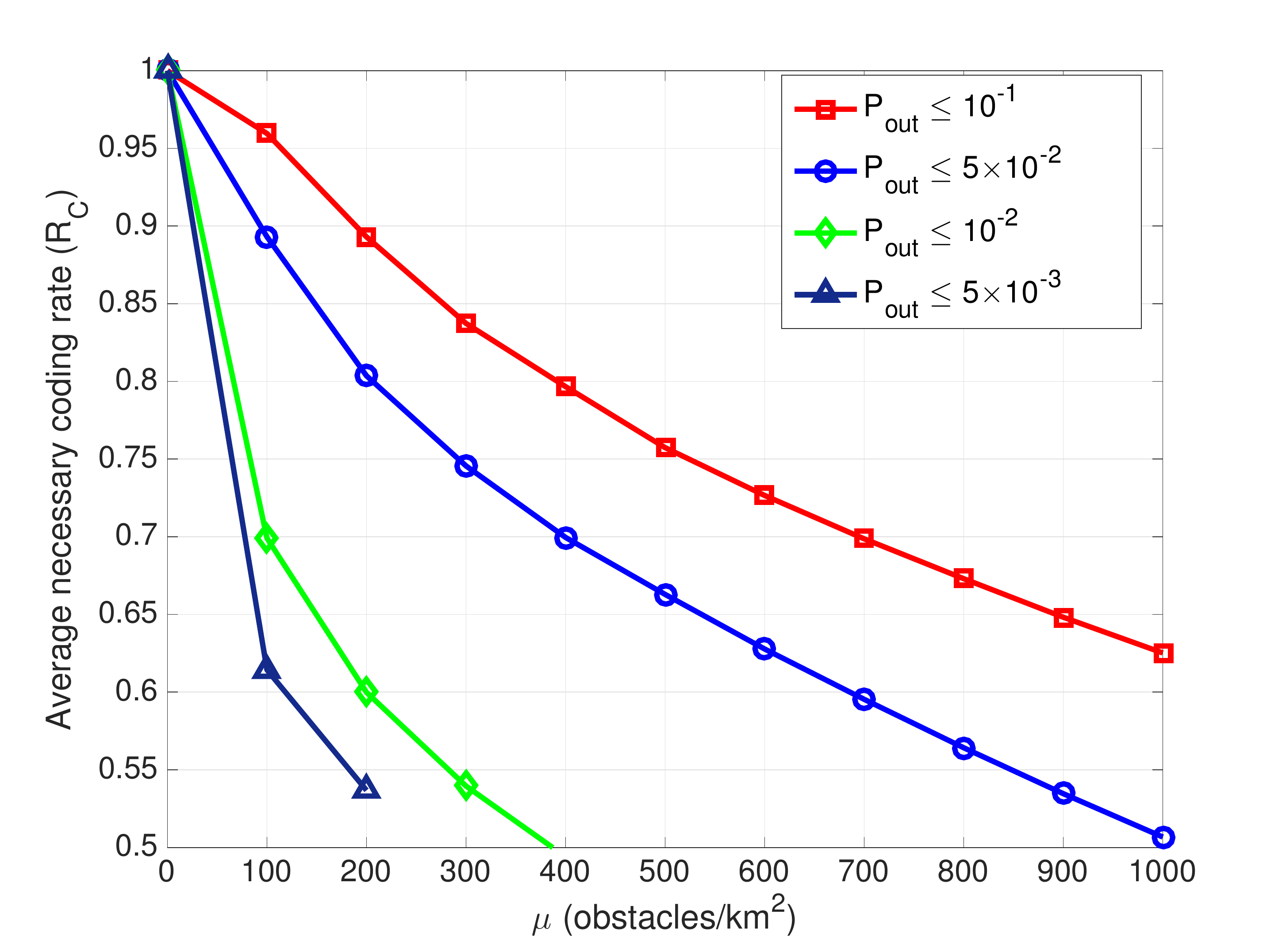}
\caption{The maximum allowed coding rate needed to guarantee that the outage probability is smaller than a given fixed value.
\label{fig:Coding_rate_vs_density}}
\end{figure}

In Fig.~\ref{fig:outage_vs_density}, we show the behaviour of the outage probability as a function of the obstacle density $\mu$. This result is obtained with the blocking probability model described in~\eqref{eq:p_on}, with $\mathbb{E}[X] = 2$ m and $\mathbb{E}[W] = 2$ m. The outage probability is computed by exhaustive evaluation of~\eqref{eq:p_out} for $R_{\min} =8$ and for all possible erasure patterns $\mathbf{e}$; the outage probability is averaged over random realizations of a deployment with $N=15$ APs randomly distributed in a square region of size $300$ m. For every deployment and for every fixed $R_{\mathcal{C}}$, the power-optimal number of links used for offloading is chosen as suggested by Corollary~\ref{cor:N_star}. The values of $\mathbb{E}[X]$, $\mathbb{E}[W]$, and $R_{\min}$ will remain constant for all the simulation results, unless stated otherwise. 
As expected, the outage probability decreases with $R_{\mathcal{C}}$ and grows with $\mu$. Fig.~\ref{fig:Coding_rate_vs_density} is obtained with the same simulation parameters of Fig.~\ref{fig:outage_vs_density}, but its goal is to show the maximum possible coding rate necessary to maintain the outage probability smaller than a fixed value. As the intuition suggests, $R_{\mathcal{C}}$ needs to decrease when $\mu$ increases, if we want to guarantee a bounded outage probability.

\subsection{To Code or Not to Code?}
This subsection addresses the following question: assuming that optimal codes can be designed for the asymmetric block-erasure channel, whose word error probability achieves the outage probability, in what circumstances are they worth to be used for power- and latency-constrained computation offloading? Some considerations and numerical simulations are provided in the sequel.

The offloading scenarios treated in this work are always conditioned by a latency constraint~\eqref{eq:R_min}. As seen in~\eqref{eq:n_i} for the uncoded case, the transmission of $n_i$ bits over the $i$-th link is power-wise optimal when it happens at a rate $R_i$ such that $n_i/n_b = R_i/R_{\min}$ and $\sum_{i=1}^NR_i = R_{\min}$. Consequently, the uplink transmission time over each channel equals $n_b/R_{\min}$ and does not depend on the number of used links. Employing a code of rate $R_{\mathcal{C}}$ to fight blocking over $N$ links implies an increase in the number of transmitted bits of a factor $R_{\mathcal{C}}^{-1}$: if $n_b$ information bits are sent in the uncoded case, they become $n_c = n_b R_{\mathcal{C}}^{-1}$ after encoding with rate $R_{\mathcal{C}}$. In this case, 
if we consider the encoding and decoding time negligible, the latency condition yields the following equivalent of~\eqref{eq:R_min}:
\begin{equation*}
  R \geq R_{\min}R_{\mathcal{C}}^{-1} =: R_{\min}'.
\end{equation*}
In other words, to keep meeting the latency constraint and at the same time send more bits over the channel(s), we need to increase our minimum transmission rate from $R_{\min}$ to $R_{\min}' \geq R_{\min}$. Consequently, the power-wise optimal number of links to exploit given by Corollary~\ref{cor:N_star} with $R_{\min}'$ replaced by $R_{\min}$ will be some $N^*_{\cod} \geq N^*$ and the inequality is generally strict. Notice also that even if $N^*_{\cod} = N^*$ (this happens, for example, when $R_{\mathcal{C}}$ is close to $1$), the tranmission rate of the coded case over the $i$-th channel $R_i'$ will \emph{not} equal $R_iR_{\mathcal{C}}^{-1}$, because the $i$-th rate in~\eqref{eq:rates} is not directly proportional to $R_{\min}$ (although linear in it). Moreover, applying~\eqref{eq:n_i}, the number of bits to be sent over the $i$-th link in the case of coded transmission will be
\begin{equation*}
  n_i' = \frac{n_b R_{\mathcal{C}}^{-1} R_i'}{R_{\min}'} = \frac{n_b R_i'}{R_{\min}}.
\end{equation*} 
Clearly, $n_i'$ is in general different from the $n_i$ of the uncoded transmission, even when $N^*_{\cod} = N^*$. The transmission time, instead, does not change: $n_i'/R_i' = n_b/R_{\min}$, coherently with the goal of meeting the same latency constraint for both the uncoded and the coded transmission scheme. When the error-correcting code is well-designed, this setup achieves the main goal of allowing the loss of information on some links (due to long-term blocking events), without compromising the offloading procedure. However, the need to transmit more bits clearly yields a cost in terms of transmission power. 
Let us call $p(R_1,\ldots,R_{N^*})$ the optimal transmission power of the uncoded scheme and $p(R_1',\ldots,R_{N_{\cod}^*}')$ the optimal transmission power of the coded scheme; they are both computed according to~\eqref{eq:total_power}, but using respectively $R_{\min}$ and $R_{\min}'$. Under what conditions $p(R_1,\ldots,R_{N^*}) > p(R_1',\ldots,R_{N_{\cod}^*}')$? Unfortunately, but not surprisingly, the answer is: under no conditions. Formally:
\begin{lemma}
  For every code $\mathcal{C}$ of rate $0 < R_{\mathcal{C}} < 1$, 
  \begin{equation*}
    p(R_1,\ldots,R_{N^*}) < p(R_1',\ldots,R_{N_{\cod}^*}').
  \end{equation*}
\end{lemma}
\begin{IEEEproof}
  We clearly see from~\eqref{eq:total_power} that, for fixed $N$, the transmission power is an increasing function of $R_{\min}$. Therefore, the power required to transmit $n_c$ coded bits at rate $R_{\min}' > R_{\min}$ over $N_{\cod}^*$ channels is always greater than the minimum power $p(R_1,\ldots,R_{N_{\cod}^*})$ required to transmit the uncoded information over the same number of channels at rate $R_{\min}$. Hence,
  \begin{equation*}
    p(R_1',\ldots,R_{N_{\cod}^*}') > p(R_1,\ldots,R_{N_{\cod}^*}) \geq p(R_1,\ldots,R_{N^*}), 
  \end{equation*}
  because transmitting over $N^*$ links is power-wise optimal in the uncoded scenario.
\end{IEEEproof}
As reasonably expectable, the previous lemma states that it is not possible to design a multi-link coded communication scheme that requires less transmission power than the corresponding optimal uncoded scheme under the same latency constraint. 
Now, in the uncoded scenario, the outage probability equals the probability that at least one link is blocked and the information sent over it is lost. Therefore, over $N$ channels, the outage probability of the uncoded transmission is $P_{\out}^{\unc}(N) = 1 - \prod_{i=1}^N(1-P_i)$. Notice that $P_{\out}^{\unc}(N)$ is a strictly increasing function of $N$, because for every $i$,
\begin{equation}
\begin{split}
\label{eq:Pout_increasing}
  P_{\out}^{\unc}(i) > P_{\out}^{\unc}(i-1) & \Leftrightarrow \prod_{j=1}^{i-1}(1-P_j) > \prod_{j=1}^{i}(1-P_j) \\
  & \Leftrightarrow 1 > 1 - P_i,
\end{split}
\end{equation}
and the latter is always true. Hence, when we restrict ourselves to the uncoded transmission scheme, we face two completely opposite requirements: the necessity to keep low (ideally to $1$) the number of channels to control the outage probability and the need for increasing it (up to $N^*$) to minimize the transmit power. We will show through numerical results in what terms coding for the block-erasure channel provides beneficial compromises between the two previous contrasting requisites. In this perspective, we claim that a fair assessment of the advantages of error-correcting codes in this scenario needs to consider the tradeoff between transmit power consumption and achievable outage probability, rather than focusing on each of these two separately. 

\begin{figure}
\includegraphics[width=\columnwidth]{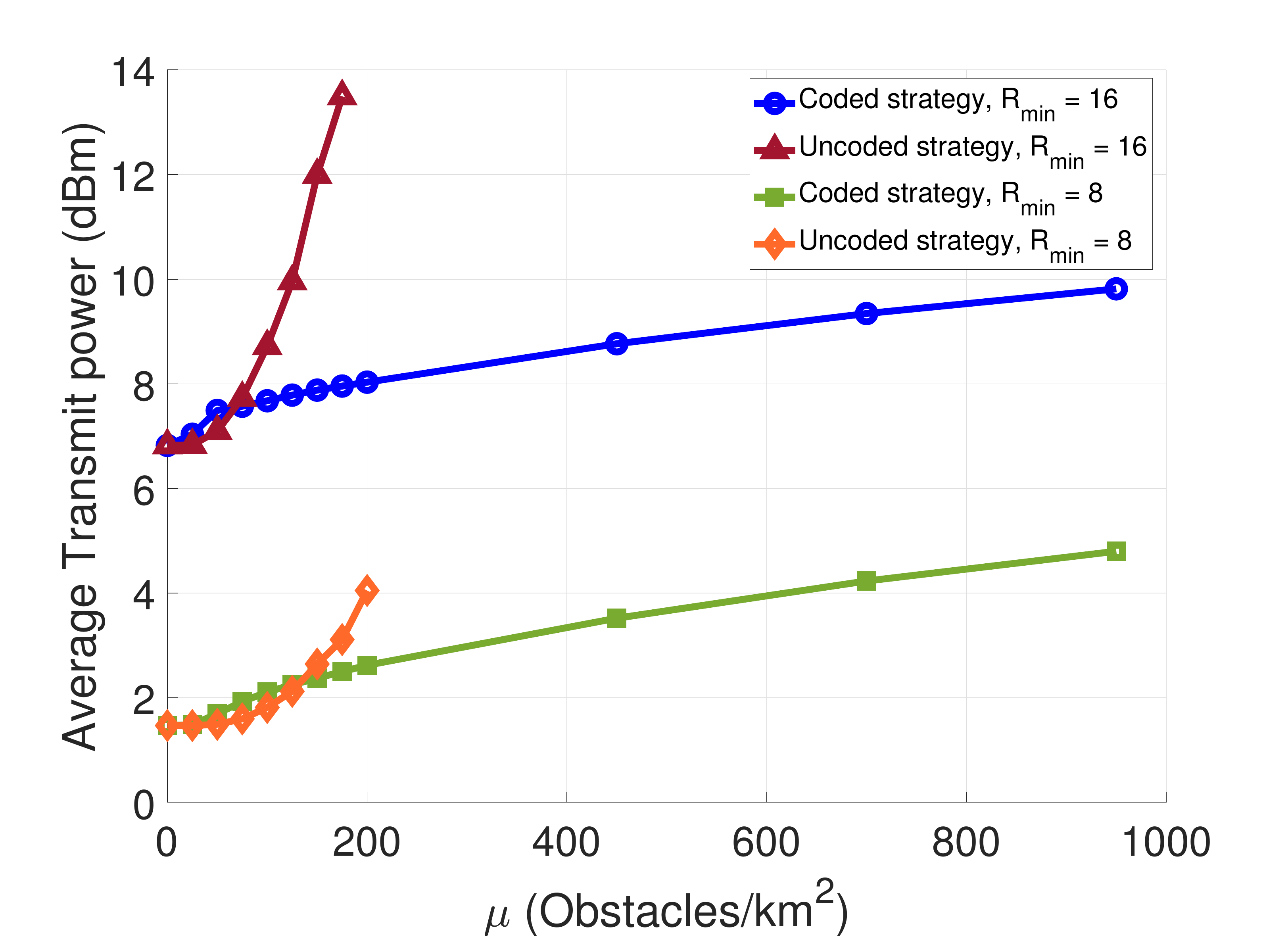}
\caption{Average transmit power in the uncoded and coded case under the constraint $P_{\out} \leq 0.05$.}
\label{fig:power_vs_mu}
\end{figure}

\begin{figure*}
\centering
\subfloat[]{\includegraphics[width=0.9\columnwidth]{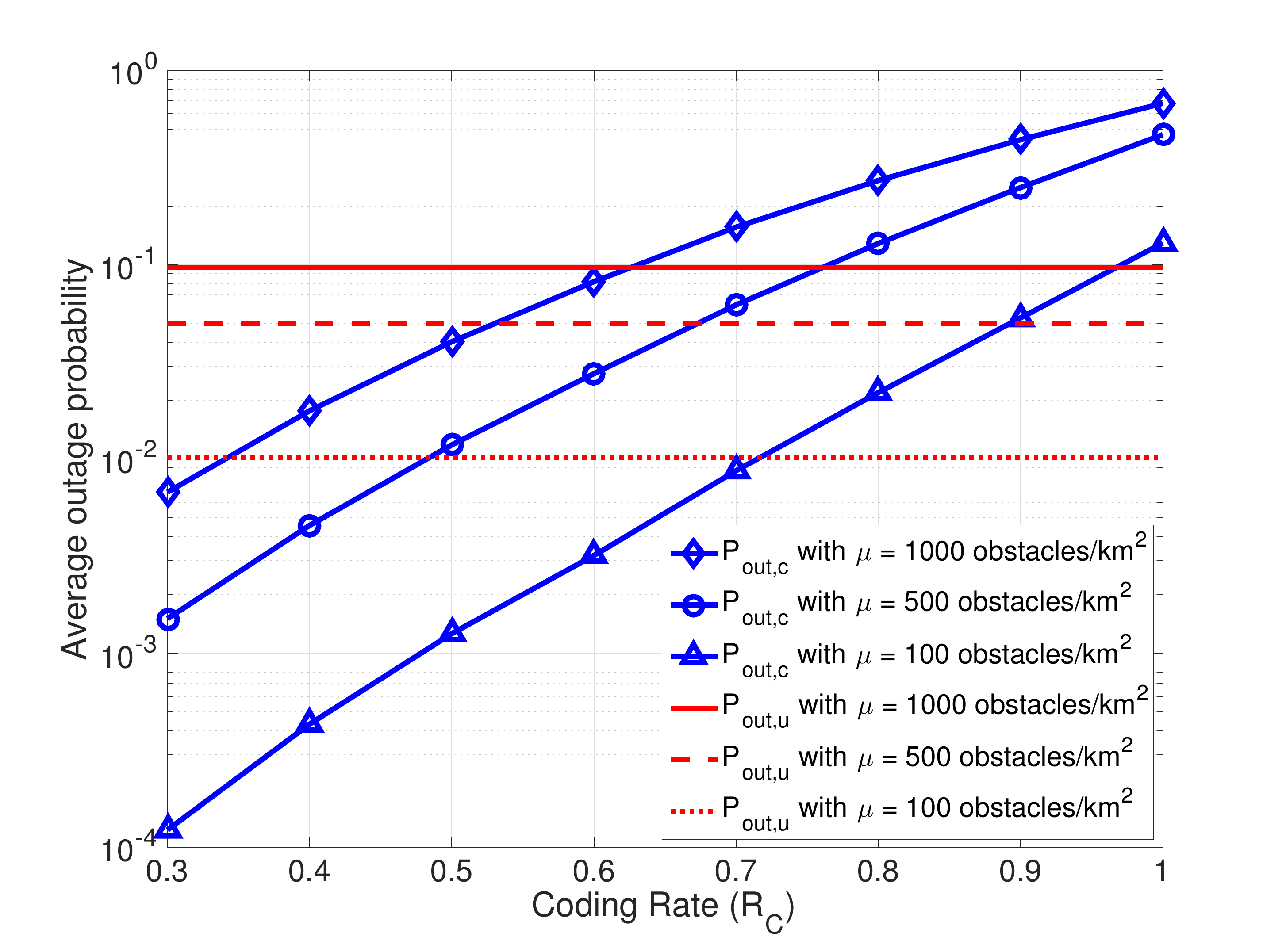}
\label{fig:Pout_vs_Rc}}
\hfil
\subfloat[]{\includegraphics[width=0.9\columnwidth]{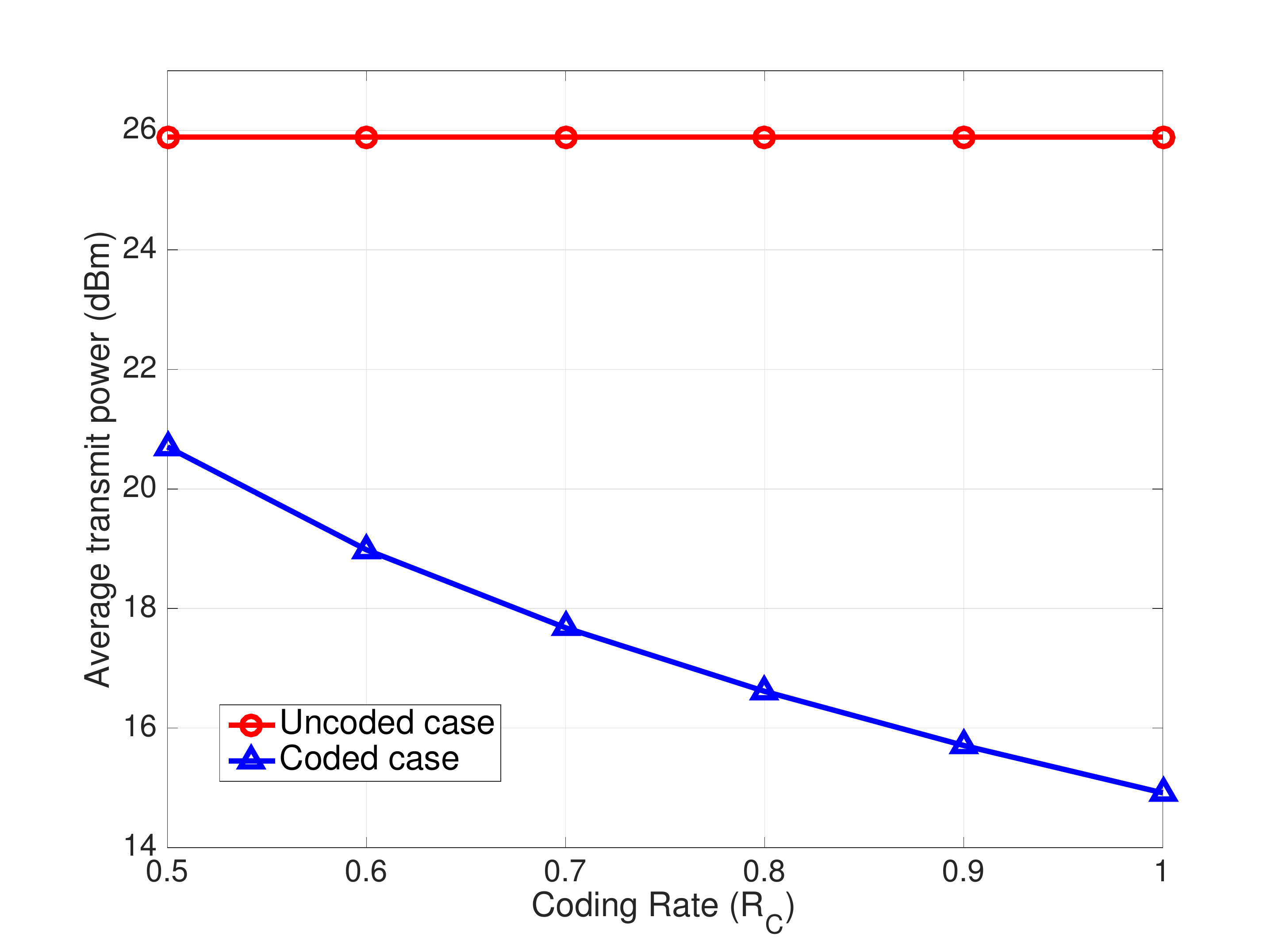}
\label{fig:Power_vs_Rc}}
\caption{Outage probability and average transmit power as functions of the coding rate $R_{\mathcal{C}}$ for different densities of obstacles.}
\label{fig:Pout_Power_vs_Rc}
\end{figure*}

Fig.~\ref{fig:power_vs_mu} 
shows the average transmit power as a function of the density of obstacles, when the outage probability is constrained below a maximum value ($P_{\out} \leq 0.05$). The results are obtained in a scenario with $15$ APs deployed in a square region of size $200$ m around the UE, where the obstacles' average dimensions are $\mathbb{E}[W]=1$ m and $\mathbb{E}[X]=2$ m. First of all, notice that if we rely on the uncoded transmission strategy, the upper bound on the outage probability can be guaranteed only for obstacle densities $\mu$ not much bigger than $175/\text{km}^2$. For higher densities, there always exist deployments in the considered region such that $P_{\out}^{\unc}(N) \geq P_{\out}^{\unc}(1) = 1-P_1 >0.05$. This is the reason why the red and orange curves in Fig.~\ref{fig:power_vs_mu} are plotted exclusively for $\mu \leq 175$. The figure depicts the comparison between the power cost of the uncoded and coded transmission strategies as a function of $\mu$ and averaged over random deployments of the $15$ APs. Recalling~\eqref{eq:Pout_increasing} and the results of Section~\ref{sec:multi_link} on transmit power minimization, the number of links $N_{\unc}$ used for uncoded multi-link offloading is computed for each instance of the AP deployment as:
  \begin{equation*}
    N_{\unc} = \max \left \{N \in \{1,\ldots,N^*\} : P_{\out} \leq 0.05 \right\} \leq N^*.
  \end{equation*}
For the coded scheme, instead, the coding rate $R_{\mathcal{C}}$ was chosen as the maximum that guarantees $P_{\out} \leq 0.05$. Then, the corresponding number of channels for multi-link offloading was computed according to Corollary~\ref{cor:N_star} with transmission rate $R_{\min}'=R_{\min}R_{\mathcal{C}}^{-1}$ and $R_{\min}=8$ or $16$. The picture clearly shows that well-designed error-correcting codes may enable offloading in scenarios where the obstacle density makes the outage probability uncontrollable for the uncoded communication strategy. Moreover, for ``medium'' obstacle densities ($75 \leq \mu \leq 175$), recurring to error-correcting codes yields considerable gains in the transmit power for $R_{\min}=16$. Finally, the figure confirms that in contexts with ``few'' obstacles (low $\mu$), a coded communication scheme may not be needed, because the outage probability remains bounded and the uncoded transmission scheme requires a smaller average transmit power.
Using the same main simulation parameters of Fig.~\ref{fig:power_vs_mu}, 
Fig.~\ref{fig:Pout_Power_vs_Rc}\subref{fig:Pout_vs_Rc} shows that error-correcting codes may also be exploited to fully outperform the best possible outage probability achievable with uncoded transmissions: the latter is obtained by exclusively transmitting over the best available link and is represented by the constant blue lines in the figure (averaged over different random AP deployments and for a few different obstacle densities in an area of $300 \text{ m} \times 300$ m). Choosing a small enough coding rate $R_{\mathcal{C}}$ allows to both obtain better average outage probabilities and to reduce the average transmit power, as shown by the combination of Fig.~\ref{fig:Pout_Power_vs_Rc}\subref{fig:Pout_vs_Rc} and Fig.~\ref{fig:Pout_Power_vs_Rc}\subref{fig:Power_vs_Rc}. For instance, an optimal code with $R_{\mathcal{C}} = 0.5$ would allow to achieve better outage probabilities than any uncoded transmission for each of the proposed obstacle densities and, at the same time, reduce by $5$ dBm the average transmit power with respect to the uncoded strategy that minimizes the outage probability.

\section{Conclusion and Directions for Future Work}
In this paper, we focused on power minimization and blocking countermeasures for computation offloading in 5G networks endowed with multi-access edge computing technologies and mmWave communication systems. First, we introduced the new paradigm of multi-link computation offloading, which relies on the capability of a user device to exploit the modern beamforming antenna technologies to generate separate simultaneous beams directed towards different mobile-edge APs. This strategy provides a new spatial degree of freedom for communications between user devices and MEC servers. In this context, we characterized the optimal solutions of the latency-constrained transmit power minimization problem for the UE both in a deterministic and a probabilistic scenario.
Then, we proposed two different methods to contrast the blocking events typical of mmWave channels: overprovisioning to compensate small information losses caused by short-term blocking and error-correcting codes for the asymmetric block-erasure channel as a solution to losses of big amounts of information due to long-term blocking events. 

Future research work on this topic may consider the possibility to relax some of our hypotheses and provide results that are based on scenarios closer to practical applications. In particular, other strategies to counteract blocking events can be investigated, such as retransmission strategies over backup links, with the necessity of introducing feedback during the communication time.
Moreover, the work on the asymmetric block-erasure channel requires further efforts for the design of close-to-optimal codes whose error probability achieves the bounds set by the outage probability. In our opinion, an interesting research topic is the design of codes that jointly protect communications against the Gaussian noise of each communication link and the block erasures that affect them.


%



\ifCLASSOPTIONcompsoc
  \section*{Acknowledgments}
\else
  \section*{Acknowledgment}
\fi
The research leading to these results is jointly funded by the European Commission H2020 and the Ministry of Internal affairs and Communications in Japan under grant agreements No 723171 5G MiEdge.

\ifCLASSOPTIONcaptionsoff
  \newpage
\fi

\end{document}